\documentclass[a4paper,11pt]{article}
\usepackage{jheppub}
\usepackage[title]{appendix}
\usepackage[hang,flushmargin]{footmisc}
\usepackage{graphicx}
\usepackage{bm}
\usepackage{amsmath}
\usepackage{amssymb}
\usepackage{amsfonts}
\usepackage{amsthm}
\usepackage{array}
\usepackage{mathtools}
\usepackage{physics}
\usepackage{hyperref}
\usepackage{orcidlink}
\usepackage{braket}
\usepackage[thinc]{esdiff}
\usepackage{wrapfig}
\usepackage{tikz}
\usepackage{enumitem}
\usepackage[framemethod=tikz]{mdframed}
\usepackage[english]{babel}
\usepackage[autostyle, english = american]{csquotes}
\usepackage[colorinlistoftodos]{todonotes}

\allowdisplaybreaks
\MakeOuterQuote{"}
\def\vev#1{\langle{#1}\rangle}

\hypersetup{colorlinks=true,urlcolor=[rgb]{0,0,0.75},citecolor=[rgb]{0.75,0,0},linkcolor=[rgb]{0,0,0.75}}

\title{\boldmath Stress-Energy Tensor of a Scalar Field on a Product Spacetime with a Time-Dependent Compact Dimension}

\author[a]{Anamitra Paul \orcidlink{0009-0000-9004-2172}}
\author[b]{, Sonia Paban \orcidlink{0000-0002-8002-842X}}
\affiliation[a]{Department of Physics, The University of Texas at Austin,\\
Austin, TX 78712, USA}
\affiliation[b]{Department of Physics, Harvard University,\\
Cambridge, MA 02138, USA}

\emailAdd{anamitra.paul@austin.utexas.edu}
\emailAdd{sppaban@fas.harvard.edu}

\abstract{We compute the vacuum expectation value of the stress-energy tensor of a scalar field on a product spacetime composed of an FLRW background times a compact dimension ($\mathcal{M}^{1, \,d-1} \times \mathcal{S}^1$), where the size of the latter is allowed to vary with time. We modify the standard adiabatic regularization prescription to obtain analytic expressions for both $d=3$ and $d=4$. In the massless and conformally coupled limit, the leading order time-dependent results are consistent with known time-independent Casimir contributions. Furthermore, in this limit, the higher-order time-dependent corrections, when the FLRW and compact-dimension scale factors coincide, match known results for ($1+d$)-dimensional FLRW spacetime.}

\begin{document}
\maketitle
\flushbottom

\section{Introduction} \label{sec:Introduction}
Within string theory, a realistic description of nature requires, in addition to the observed four-dimensional Friedmann–Lemaître–Robertson–Walker (FLRW) spacetime, the presence of additional small, compact spatial dimensions. These extra dimensions are generically unstable: a natural tendency in string theory is for them to expand and decompactify \cite{Dine:1985he}. Consequently, a great deal of work has focused on identifying stabilization mechanisms; see \cite{McAllister:2023vgy} for a recent review. Among the various contributions to the effective potential that could stabilize the internal dimensions at a finite size is the Casimir energy. There is an extensive literature computing the Casimir energy in Kaluza–Klein theories \cite{Appelquist:1982zs, Appelquist:1983vs, Appelquist:1983wx, Weinberg:1983xy, Candelas:1983ae, Kantowski:1986iv,Elizalde:1988rh} and, more recently, in string-theoretic contexts \cite{DeLuca:2021pej, ValeixoBento:2025yhz, DallAgata:2025jii}. Almost all of these studies assume that the size of the extra dimensions is time-independent, primarily because their objective is to identify static stable vacua. There are also strong experimental constraints on any time variation in the size of the extra dimensions, since such variation would induce changes in observable fundamental constants. Atomic clock experiments constrain the current variation of the fine structure constant, $|\dot{\alpha}/\alpha_0| < 10^{-17} \rm{yr}^{-1}$ \cite{Rosenband:2008qgq}, while ALMA measurements of absorption spectroscopy toward high-redshift quasars bound the variation $\Delta \alpha= \alpha(z=0.89)-\alpha_0$ to $|\Delta{\alpha}/\alpha_0| < 0.55 \rm{p.p.m}$ \cite{Muller:2025azz}, which roughly translates to $|\dot{\alpha}/\alpha_0| < 10^{-16} \rm{yr}^{-1}$, under the assumption of a constant rate. Similarly, there are constraints on the time variation of Newton's gravitational constant: local measurements using Lunar Laser Ranging and globular clusters imply $|\dot{G}/G_0| \lesssim 10^{-14} \rm{yr}^{-1}$, and Big Bang nucleosynthesis (BBN) bounds, assuming constant $\dot G$, give $|\dot{G}/G_0| \lesssim 10^{-12} \rm{yr}^{-1}$ \cite{Uzan:2024ded, Martins:2017yxk}. For comparison, $ H_0= |\dot{a}/{a}| \sim 7 \times 10^{-11} \rm{yr}^{-1}$. This raises the question of whether time-dependent corrections to the Casimir energy can play a significant role at any stage in the history of the universe. Given the experimental bounds, such effects, if important, are most likely to be relevant only in the early universe.

The correction to the Casimir energy due to moving boundaries, the dynamical Casimir effect, was first observed in a superconducting circuit in 2011 \cite{Wilson:2011rsw}, in a context where gravity can be neglected. In cosmology, although quite a bit of work has been devoted to compactification dynamics (for examples, see \cite{Chodos:1979vk,Freund:1982pg,Alvarez:1983kt,Yoshimura:1984qa,Kolb:1986nj, Giddings:2004vr, Carroll:2009dn, Chen:2005ae,Dasgupta:2018rtp,Dasgupta:2014pma,Kamal:2025qia}), we have found only one reference \cite{Maeda:1984un} that has attempted to compute the time dependent corrections when gravity matters. A complete calculation of the time-dependent Casimir effect has not been performed; even in the simplest scenario—an FLRW background with a compact dimension ($\mathcal{M}^{1, \, d-1} \times \mathcal{S}^1$)—there is no well-established practical method for eliminating the UV divergences. The adiabatic regularization procedure introduced by Parker and Fulling \cite{Parker:1974qw, Fulling:1974pu}, though well suited to cosmology, has only been shown to be equivalent to point-splitting regularization when the scalar field is massless and conformally coupled, while the spacetime $\mathcal{M}^{1, \, d-1} \times \mathcal{S}^1$ is conformally flat i.e. when the scale factors of the two product spaces are equal \cite{Birrell:1978rs}. This limitation arises because adiabatic regularization requires exact solutions for the field modes, which are not known for a generic field mass and Ricci coupling, or a general $\mathcal{M}^{1, \, d-1} \times \mathcal{S}^1$ compactification. To address this challenge and because we are physically interested in the adiabatic limit, we propose a modification to the standard adiabatic regularization prescription, which we explain in section~\ref{sec:Tformalconstruction}. In summary, rather than employing the precise field modes, this approach utilizes a WKB approximation to determine the modes. For a massless and conformally coupled field on a conformally flat background, the exact mode solutions are known; in this case, the expectation of the stress-energy obtained using the modified prescription agrees with that derived using the standard prescription. Consequently, although a direct comparison between the approximate prescription and point-splitting is not conducted in the general case, there is concordance between them in the above limit, since point-splitting agrees with the standard adiabatic prescription. Outside this limit, we require that the stress-energy tensor be conserved and that the outcomes reproduce the conformally flat case. The method is checked for the cases $d=3$ and $d=4$. In both instances, the regularized stress-energy tensor is finite and its covariant divergence vanishes. Furthermore, it reproduces standard FLRW results when the scale factors are identical and the field is massless and conformally coupled. The calculation for the case $d=3$ was performed not for physical relevance but to further verify the proposed regularization procedure, given that we anticipate a non-zero trace of the stress-energy tensor within the FLRW limit \cite{Wald:1978pj, Birrell:1982ix, Parker:2009uva}.

The goal of this work is to compute the time-dependent corrections to the Casimir energy/pressures and to test the proposed regularization procedure. The cosmological implications are deferred to later work. However, as shown in section~\ref{sec:EinsteinFrame}, a comparison between the first adiabatic corrections reveals that these corrections, in $1+4$ dimensions, will be significant during early times, specifically when the product $r_0 M_{5} b \ll 1$ (here $r_0$ is the compact dimension’s scale, $M_{5}$ is the reduced Planck mass in $1+4$ dimensions, and $b$ is the internal scale factor). In $1+3$ dimensions, the first non-trivial corrections emerge at $4$-th adiabatic order, resulting in equations with higher derivatives.\footnote{The presence of higher derivatives in the equations of motion has the potential to alter the time evolution significantly \cite{Simon:1990ic}.} A realistic cosmological analysis must incorporate the contributions of fermions and other higher-spin particles, which have not been considered in the present study as the primary objective is to verify the proposed regularization procedure. The computation we present assumes that the scalars are free and non-minimally coupled to gravity. A more realistic scenario would include additional contributions to the potential beyond the Casimir energy and the non-minimal scalar coupling to gravity. Retaining the latter is equivalent to choosing a specific potential in the Einstein frame. Despite this arbitrariness, we retain it to verify the approximation, as it appears in many of the preceding limits to which we compare our results. We do not know how to find the mode decomposition for an arbitrary potential. Estimating the Casimir energy in the general case might be possible only within perturbation theory. 

The rest of the paper is organized as follows. In section~\ref{sec:Setup}, we present the setup to compute $\vev{T_{A B}}$. In section~\ref{sec:Adiabatic Regularization}, we review the adiabatic regularization procedure and explain the proposed modification to the standard prescription. In section~\ref{sec:Tregcomputation}, we explicitly carry out the computation of the regularized stress-energy tensor for generic field mass and Ricci coupling. In section~\ref{sec:MasslessConformalLimit}, we analyze the massless and conformally coupled limit for both $d=3$ and $d=4$ to compare with known results in the literature. Finally, in section~\ref{sec:EinsteinFrame}, we construct the dimensionally reduced Einstein frame action for the $d=4$ scenario and demonstrate how to deal with divergences that arise in the Jordan frame stress-energy computation.

\section{Setup} \label{sec:Setup}
We start with the action
\begin{align}
    S = \int d^{1+d} x \, |g|^{1/2} \left[ \frac{M_{1+d}^{d-1}}{2} \mathcal{R}_{(1+d)} +  \frac{1}{2} g^{A B} \nabla_A \Phi \, \nabla_B \Phi - \frac{1}{2} \left( m^2 + \xi \,\mathcal{R}_{(1+d)} \right) \Phi^2  \right]
    \label{eq:d_action}
\end{align}
where $M_{1+d}$ is the reduced Planck mass and $\mathcal{R}_{(1+d)} = \mathcal{R}\left[g_{AB}\right]$ is the Ricci scalar in $1+d$ dimensions. Our sign conventions\footnote{Particularly, we use $\mathcal{R}^{A}{}_{B C D} = \partial_{D}\Gamma^{A}{}_{B C} - \partial_{C}\Gamma^{A}{}_{B D} + \Gamma^{M}{}_{B C}\Gamma^{A}{}_{D M} - \Gamma^{M}{}_{B D}\Gamma^{A}{}_{C M} \hspace{1mm}$ and $\mathcal{R}_{A B} = \mathcal{R}^{C}{}_{A C B}\,$.} for the metric, curvature tensors, and Einstein field equations are $(---)$ in the terminology of Misner et al. \cite{Misner:1973prb}. Let us assume a solution for the metric of the form
\begin{align}
    ds^2 = g_{A B} \, dx^A \,dx^B = dt^2 - a(t)^2 \sum_{j=1}^{d-1} dx_j^2- b(t)^2 dx_{d}^2 \notag\\
    x_i \in \left(-\infty, \infty \right), \; x_{d} \in \left[ 0, 2 \pi r_0 \right], \; i \in \left\{ 1, ..., d-1 \right\}
    \label{eq:d_FLRW}
\end{align}
The equation of motion is
\begin{align}
    \left( \Box + m^2 + \xi \, \mathcal{R}_{(1+d)} \right) \Phi = 0
    \label{eq:genericEOM}
\end{align}
which using $\left( \ref{eq:d_FLRW} \right)$ gives 
\begin{align}
    \partial_t^2 \Phi 
    &- \frac{1}{a^2} \sum_{i=1}^{d-1} \partial_i^2 \Phi 
    - \frac{1}{b^2} \partial_d^2 \Phi 
    + \left( \frac{\left(d - 1\right) \dot a}{a} + \frac{\dot b}{b} \right) \partial_t \Phi 
    + m^2 \Phi \notag\\
    &+ \frac{\left( d - 1 \right) \left( d - 2 \right) \, \xi \, \dot a^2}{a^2} \Phi + 2 \left(d-1\right) \, \xi \left(\frac{\ddot a}{a} +\frac{\dot a \dot b}{a b}\right) \Phi 
    + \frac{2 \, \xi \, \ddot b}{b} \Phi = 0
\end{align}
The field may be expanded with a set of modes
\begin{align}
    \Phi(x) = \sum_{n=-\infty}^{\infty} \int_{-\infty}^{\infty} d^{d-1} k \left[ A_{nk} f_{nk} \left( x \right) + A_{nk}^\dagger f_{nk}^* \left( x \right) \right]
    \label{eq:modedecomp}
\end{align}
that are orthonormal with respect to the scalar product
\begin{align}
    \left( f_1, f_2 \right) = i \int d^{d} x \, |g|^{1/2} \, g^{0 A} f_1^*(t, \vec{x}) \ \overset{\leftrightarrow}{\partial_A} \, f_2(t, \vec{x})
    \label{eq:kgscalarprod}
\end{align}
The canonical commutation relations are given by
\begin{align}
    \left[ A_{nk}, A_{n'k'}^\dagger \right] = \delta_{kk'} \delta_{nn'}\,, \hspace{1cm} \left[ A_{nk}, A_{n'k'} \right] = 0
\end{align}
The stress-energy tensor is
\begin{align}
    T_{A B} =& \ \nabla_A \Phi \nabla_B \Phi - \frac{1}{2} g_{A B} \, g^{C D} \, \nabla_C \Phi \nabla_D \Phi + \frac{1}{2} g_{A B} \, m^2 \Phi^2 \notag\\
    &- \xi \left\{ \mathcal{R}_{(1+d)AB} - \frac{1}{2} \mathcal{R}_{(1+d)} \, g_{A B} \right\} \Phi^2 + \xi \left\{ g_{A B} \, \Box \left( \Phi^2 \right) - \nabla_A \nabla_B \left( \Phi^2 \right) \right\}
\end{align}
Plugging in the mode decomposition \eqref{eq:modedecomp}, we get the vacuum expectation
\begin{align}
    \bra{0} T_{A B} \ket{0} =& \: \sum_{n=-\infty}^{\infty} \int_{-\infty}^{\infty} d^{d-1} k \biggl[ \nabla_A f_{nk} \nabla_B f_{nk}^* - \frac{1}{2} g_{A B} \, g^{C D} \, \nabla_C f_{nk} \nabla_D f_{nk}^* \notag \\
    &+ \frac{1}{2} g_{A B} \, m^2 f_{nk} f_{nk}^* - \xi \left\{ \mathcal{R}_{(1+d)AB} - \frac{1}{2} \mathcal{R}_{(1+d)} \, g_{A B} \right\} f_{nk} f_{nk}^* \notag\\
    &+ \xi \left\{ g_{A B} \, \Box \left( f_{nk} f_{nk}^* \right) - \nabla_A \nabla_B \left( f_{nk} f_{nk}^* \right) \right\} \biggl]  \label{eq:genericTVEV}
\end{align}

\section{Adiabatic regularization} \label{sec:Adiabatic Regularization}
The formal expression \eqref{eq:genericTVEV} is filled with UV divergences. Adiabatic regularization, a method to systematically remove such divergences in homogeneous spacetimes, was first introduced in \cite{Parker:1974qw, Fulling:1974pu} and shown to be equivalent to point-splitting regularization in \cite{Birrell:1978rs}. For the physically relevant limit of a slowly varying spacetime, no particle production should take place. Orthonormal field mode solutions are chosen such that the lowest order terms in their adiabatic expansions (in the number of time derivatives) match Minkowski modes at early (or late) times, a requirement known as the adiabatic condition. This condition ensures that the above constraint on particle number is satisfied. A set of modes satisfying the adiabatic condition define an adiabatic vacuum. While the condition does not necessarily yield a unique set of modes (and hence vacuum), the large momentum behavior of physical observables are the same, regardless of which specific set is chosen. Thus, the leading order quantities constructed using the adiabatic expansion of the modes must be subtracted from the formal expression of the observable constructed with the exact field modes to yield a finite result. A detailed discussion of this is provided in \cite{Parker:2009uva}.

Furthermore, one expects a good renormalization scheme to maximally subtract UV divergences, while leaving the IR behavior unchanged. The subtracted quantities in adiabatic regularization should depend only on the local behavior, not the global topology of the spacetime. Most applications of adiabatic regularization involve computing observables such as the stress-energy tensor in open spacetimes. In such cases, both the formal expression and the adiabatic expression to be subtracted are constructed from mode solutions using a continuum measure. However, for closed spacetimes of interest such as ours, expressions for the stress-energy tensor involve a discrete sum over mode solutions, as in \eqref{eq:genericTVEV}. It has been argued that in such situations, the proper quantity to subtract using the adiabatic method should still be constructed from mode solutions with a continuum measure instead of a discrete one \cite{Anderson:1987yt}.

\subsection{Constructing \texorpdfstring{$\vev{T_{A B}}_{\rm{A}}$}{}} \label{sec:Tadbconstruction}
Let us make the above ideas more precise within the context of our scenario. We want to find the quantity $\vev{T_{A B}}_{\rm{A}}$ that must be subtracted according to adiabatic regularization from the formal expression of the stress-energy tensor to obtain
\begin{align}
    \vev{T_{A B}}_{\rm{reg}} : \! &= \vev{T_{A B}} - \vev{T_{A B}}_{\rm{A}}
    \label{eq:regVEVgeneric}
\end{align}
Keeping in mind that the quantity to be subtracted must be constructed with a continuum measure, consider the metric
\begin{align}
    ds^2 = dt^2 - a(t)^2 \sum_{j=1}^{d-1} dx_j^2- b(t)^2 dx_{d}^2 \,, \hspace{1cm} x_i \in \left(-\infty, \infty \right), \; i \in \left\{ 1, ..., d \right\} \label{eq:d_FLRWNC}
\end{align}
It is similar to \eqref{eq:d_FLRW}, but the $d$-th spatial dimension is not compactified. However, the local curvature tensors are the same for both metrics. For the non-compact spacetime, the field may be decomposed into orthonormal mode solutions of \eqref{eq:genericEOM} as
\begin{align}
    \Phi(x) = \int_{-\infty}^{\infty} d^{d} k \left[ A_{k} f_{k} \left( x \right) + A_{k}^\dagger f_{k}^* \left( x \right) \right]
    \label{eq:ncmodedecomp}
\end{align}
where 
\begin{align}
    f_{k}(x) = \left( 2 ( 2 \pi )^{d} \right)^{-1/2} a(t)^{-(d-1)/2} b(t)^{-1/2} e^{i \left( \vec{k} \cdot \vec{x} + k_{d} x_{d} \right)} \, h_{k}(t)
    \label{eq:ncfmode}
\end{align}
For orthonormality, the condition
\begin{align}
    h_{k} \, \dot h_{k}^* - h_{k}^* \, \dot h_{k} = 2i
\end{align}
must be satisfied. Plugging in \eqref{eq:ncfmode} into \eqref{eq:genericEOM}, gives
\begin{align}
    \frac{d^2}{dt^2} h_{k} + \left( \omega_{k}^2 + \sigma \right) h_{k} = 0
    \label{eq:nchdiffeq}
\end{align}
where
\begin{align}
    \omega_{k} = \left( \sum_{i=1}^{d-1} \frac{k_{i}^2}{a^2} + \frac{k_{d}^2}{b^2} + m^2 \right)^{1/2} = \left( \frac{k_{T}^2}{a^2} + \frac{k_{d}^2}{b^2} + m^2 \right)^{1/2} \label{eq:NComega}
\end{align}
\begin{align}
    \sigma = \frac{\left(1-d\right) \! \left(1-4 \xi \right)}{2} \left(\frac{\ddot a}{a}+\frac{\dot a \dot b}{a b}\right) -\frac{1}{4} \left(1-d\right) \! \left(1+\left(2-d\right)\left(1- 4 \xi\right) \right) \frac{{\dot a}^2}{a^2} + \frac{1}{4} \frac{{\dot b}^2}{b^2 } - \frac{\left(1-4 \xi \right)}{2} \frac{\ddot b}{b}
\end{align}
A solution of \eqref{eq:nchdiffeq} can be expanded up to $N$-th adiabatic order as
\begin{align}
    h_{k}(t) &= W_{k}(t)^{-1/2} \exp \left( -i \int^t W_{k}(t') \, dt' \right) \label{eq:hsoladb}\\
    W_{k} &= W_k^{(0)} + W_k^{(1)} + ... + W_k^{(N)} \label{eq:WAdbExpansion}
\end{align}
For large $k$, \eqref{eq:hsoladb} asymptotes to Minkowski modes, satisfying the adiabatic condition mentioned previously. As shown in \cite{Parker:2009uva}, each term on the right hand side of \eqref{eq:WAdbExpansion} can be obtained by plugging it into
\begin{align}
    W_{k}^2 = \omega_{k}^2 + \sigma + W_{k}^{1/2} \frac{d^2}{dt^2} W_{k}^{-1/2} \label{eq:adbdiffeq}
\end{align}
and equating terms of the same order on both sides. This gives us
\begin{align}
    W_{k}^{(0)} &= \omega_{k} \label{eq:W0}\\
    W_{k}^{(1)} &= W_{k}^{(3)} = W_{k}^{(5)} = 0\\
    W^{(2)} &= -\frac{\ddot \omega_{k}}{4 \omega_{k}^2}+\frac{3 \dot \omega_{k}^2}{8 \omega_{k}^3}+\frac{\sigma}{2 \omega_{k}} \label{eq:W2}\\
    W_{k}^{(4)} &= \frac{\omega_{k}^{(4)}}{16 \omega_{k}^4}+\frac{3 \sigma \ddot \omega_{k}}{8 \omega_{k}^4}-\frac{13 \ddot \omega_{k}^2}{32 \omega_{k}^5}+\frac{5 \dot \omega_{k} \dot \sigma}{8 \omega_{k}^4}-\frac{19 \sigma \dot \omega_{k}^2}{16 \omega_{k}^5}-\frac{297 \dot \omega_{k}^4}{128 \omega_{k}^7}-\frac{5 \omega_{k}^{(3)} \dot \omega_{k}}{8 \omega_{k}^5} \notag\\
    &\hspace{5mm}+\frac{99 \dot \omega_{k}^2 \, \ddot \omega_{k}}{32 \omega_{k}^6} -\frac{\ddot \sigma}{8 \omega_{k}^3}-\frac{\sigma^2}{8 \omega_{k}^3} \label{eq:W4}\\
    & \vdotswithin{ = }\notag
\end{align}

The regularization scheme requires that given an adiabatic expansion of a physical observable, if any term at a given order diverges, then all terms of that same order must be subtracted from the formal expression for the observable, even if some of them may be finite. Since we are considering spacetimes with $d+1$ dimensions, some terms in the adiabatic expansion of $\vev{T_{A B}}$ diverge up to ($d+1$)-th order. Our computations involved $d=3$ and $d=4$, and hence, we only need to consider terms up to $4$-th and $5$-th order respectively in the expansion \eqref{eq:WAdbExpansion}. Now that the adiabatic expansion of the non-compact modes \eqref{eq:ncfmode} are fully specified, the quantity that needs to be subtracted from the formal stress-energy expectation is
\begin{align}
    \vev{T_{A B}}_{\rm{A}} =& \int_{-\infty}^{\infty} d^{d} k \biggl[ \nabla_A f_{k} \nabla_B f_{k}^* - \frac{1}{2} g_{A B} \, g^{C D} \, \nabla_C f_{k} \nabla_D f_{k}^* + \frac{1}{2} g_{A B} \, m^2 f_{k} f_{k}^* \notag \\
    & - \xi \left\{ \mathcal{R}_{(1+d)AB} - \frac{1}{2} \mathcal{R}_{(1+d)} \, g_{A B} \right\} f_{k} f_{k}^* + \xi \left\{ g_{A B} \, \Box \left( f_{k} f_{k}^* \right) - \nabla_A \nabla_B \left( f_{k} f_{k}^* \right) \right\} \biggr] \label{eq:genericNCTVEV}
\end{align}

\subsection{Constructing \texorpdfstring{$\vev{T_{A B}}$}{} with an approximate prescription} \label{sec:Tformalconstruction}
The adiabatic method requires a set of exact field modes $f_{nk}$ of the compact spacetime of interest in order to construct the formal expression \eqref{eq:genericTVEV}. However, these field modes are unknown for generic $a(t)$ and $b(t)$, with closed-form solutions existing only for the conformally flat scenario $a(t)=b(t)$. We propose modifying the standard prescription by using a WKB type approximation (similar to the expansion of the non-compact modes) for the compact solutions instead of the exact ones.
\begin{align}
    f_{nk}(x) &\approx \left( 2 ( 2 \pi )^{d-1} \left( 2 \pi r_0 \right) \right)^{-1/2} a(t)^{-(d-1)/2} b(t)^{-1/2} e^{i \left( \vec{k} \cdot \vec{x} + \frac{n}{r_0} x_d \right) } \, h_{nk}(t) \label{eq:hsolcompapprox}\\
    h_{nk}(t) &= W_{nk}(t)^{-1/2} \exp \left( -i \int^t W_{nk}(t') \, dt' \right)
\end{align}
where $W_{nk}$ satisfies equations identical to \eqref{eq:adbdiffeq}-\eqref{eq:W4} with $\omega_k$ replaced by 
\begin{align}
    \omega_{nk} = \left( \sum_{i=1}^{d-1} \frac{k_{i}^2}{a^2} + \frac{\left( n/r_0 \right)^2}{b^2} + m^2 \right)^{1/2} = \left( \frac{k_{T}^2}{a^2} + \frac{\left( n/r_0 \right)^2}{b^2} + m^2 \right)^{1/2}
\end{align}

To evaluate the validity of our prescription, we will check that the expression for $\vev{T_{A B}}_{\rm{reg}}$ satisfies the following conditions:
\begin{enumerate}
\item $\vev{T_{A B}}_{\rm{reg}}$ is finite.
\item The stress-energy is conserved i.e. $\nabla_{A} \vev{T^{A B}}_{\rm{reg}}=0$.
\item It reproduces the known calculable results using the standard adiabatic regularization prescription when $a(t)=b(t)$.
\end{enumerate}

\section{Computing \texorpdfstring{$\bm{\vev{T_{A B}}_{\rm{reg}}}$}{}} \label{sec:Tregcomputation}

Since both the formal and adiabatic expressions for the stress-energy tensor are UV divergent, they cannot be directly subtracted from each other. We have to separately regularize them first using other techniques to make them finite and well defined. We will perform this step using a combination of dimensional and zeta regularization, similar to the computations of Elizalde and Romeo \cite{Elizalde:1988rh}. Once this step has been completed, we will carry out the adiabatic subtraction to obtain a finite value for \eqref{eq:regVEVgeneric}.

We are primarily interested in the $d=4$ case and present its method of computation (the $d=3$ case follows analogously). By symmetry, we know that $\vev{T_{00}} \neq \vev{T_{11}} = \vev{T_{22}} = \vev{T_{33}} \neq \vev{T_{44}}$ and that all the off-diagonal components are zero. Focusing on the $00$-component for a moment, let us regularize the adiabatic expression \eqref{eq:genericNCTVEV} first. Plugging in \eqref{eq:ncfmode}, we get
\begin{align}
    \vev{T_{00}}_{\rm{A}} =& \, \frac{1}{32 \pi^4 a^3 b}\int_{-\infty}^{\infty} d^4 k \biggl[ \frac{1}{2} \frac{\omega_{k}^2}{W_{k}} + \frac{1}{2} W_{k} + 
    \left( \frac{9}{8} - 6 \xi \right) \frac{\dot a^2}{a^2 W_{k}} + 
    \left( \frac{3}{4} - 3 \xi \right) \frac{\dot a \dot b}{a b W_{k}} \notag\\
    &+ \left( \frac{1}{8} - \xi \right) \frac{\dot b^2}{b ^2 W_{k}} + 
    \left( \frac{3}{4} - 3 \xi \right) \frac{\dot a \dot W_{k}}{a W_{k}^2} + 
    \left( \frac{1}{4} - \xi \right) \frac{\dot b \dot W_{k}}{b  W_{k}^2} + 
    \frac{\dot W_{k}^2}{8 W_{k}^3} \biggl]
    \label{eq:bianchiNCTVEV}
\end{align}
Using \eqref{eq:W0}-\eqref{eq:W4}, the integrand of \eqref{eq:bianchiNCTVEV} can be adiabatically expanded\footnote{The integrand expansions for the components $i = 1, 2, 3$ on the diagonal explicitly contain terms of the form $\mathcal{I} \left( k_T^2, k_i^2 \right) = h\!\left( k_T^2 \right) k_i^2$ for some function $h$. As $k_T^2 = \sum_{i=1}^3 k_i^2$ and the integral $\int d^4 k \, \mathcal{I} \left( k_T^2, k_i^2 \right)$ is invariant upon interchanging $k_1 \leftrightarrow k_2 \leftrightarrow k_3$, performing the substitution $k_{i}^2 \rightarrow \frac{1}{3} k_{T}^2$ preserves the value of the overall integral, while still letting us use the general formalism outlined in this section.} as
\begin{align}
    \mathcal{I}_{00} =& \ \mathcal{I}_{00}^{(0)} + \mathcal{I}_{00}^{(2)} + \mathcal{I}_{00}^{(4)} \label{eq:NCIntegrandExpansion}\\
    \mathcal{I}_{00}^{(0)} =& \ \omega_{k} \label{eq:NCIntegrand0}\\
    \mathcal{I}_{00}^{(2)} =& \left( \frac{9}{8} - 6 \xi \right) \frac{\dot a^2}{a^2 \omega_{k}} + 
    \left( \frac{3}{4} - 3 \xi \right) \frac{\dot a \dot b}{a b \omega_{k}} + 
    \left( \frac{1}{8} - \xi \right) \frac{\dot b^2}{b ^2 \omega_{k}} + 
    \left( \frac{3}{4} - 3 \xi \right) \frac{\dot a \dot \omega_{k}}{a \omega_{k}^2} \notag\\
    &+ \left( \frac{1}{4} - \xi \right) \frac{\dot b \dot \omega_{k}}{b  \omega_{k}^2} + \frac{1}{8} \frac{\dot \omega_{k}^2}{\omega_{k}^3} \label{eq:NCIntegrand2}\\
    \mathcal{I}_{00}^{(4)} =& \, \frac{1}{\omega_{k}^3} \Biggl[ \left(3 \xi -\frac{9}{16}\right) \frac{\dot{a}^2 \sigma }{a^2} + \left(\frac{3 \xi }{2}-\frac{3}{8}\right) \frac{\dot{a} \dot{b } \sigma }{a b} + \left(-\frac{3 \xi }{2}+\frac{3}{8}\right) \frac{\dot{a} \dot{\sigma }}{a} + \left(\frac{\xi }{2}-\frac{1}{16}\right) \frac{\dot{b }^2 \sigma }{b^2} \notag \\
    &+ \left(-\frac{\xi }{2}+\frac{1}{8}\right) \frac{\dot{b } \dot{\sigma }}{b} + \frac{\sigma ^2}{8}\Biggr] + \frac{\dot{\omega_{k} }}{\omega_{k} ^4} \Biggl[ \left(\frac{9 \xi }{2}-\frac{9}{8}\right) \frac{\dot{a} \sigma }{a} + \left(\frac{3 \xi }{2}-\frac{3}{8}\right) \frac{\dot{b } \sigma }{b} + \frac{\dot{\sigma }}{8} \Biggr] \notag \\
    &+ \frac{\dot \omega_{k}^2}{\omega_{k} ^5} \Biggl[ \left(\frac{9 \xi }{4}-\frac{27}{64}\right) \frac{\dot{a}^2}{a^2} + \left(\frac{9 \xi }{8}-\frac{9}{32}\right) \frac{\dot{a} \dot{b }}{a b} + \left(\frac{3 \xi }{8}-\frac{3}{64}\right) \frac{\dot{b }^2}{b ^2} - \frac{\sigma }{8} \Biggr] + \frac{5 \dot \omega_{k}^2 \ddot \omega_{k}}{16 \omega_{k} ^6} \notag\\
    &+ \frac{\dot \omega_{k}^3}{\omega_{k} ^6} \Biggl[ \left(\frac{45 \xi }{8}-\frac{45}{32}\right) \frac{\dot{a}}{a} 
    + \left(\frac{15 \xi }{8}-\frac{15}{32}\right) \frac{\dot{b }}{b } \Biggr] -\frac{45 \dot \omega_{k}^4}{128 \omega_{k} ^7} 
    + \frac{\ddot \omega_{k}}{\omega_{k} ^4} \Biggl[ \left(-\frac{3 \xi }{2}+\frac{9}{32}\right) \frac{\dot{a}^2}{a^2} \notag\\
    &+ \left(-\frac{3 \xi }{4}+\frac{3}{16}\right) \frac{\dot{a} \dot{b }}{a b} 
    + \left(-\frac{\xi }{4}+\frac{1}{32}\right) \frac{\dot{b }^2}{b ^2} - \frac{\sigma }{8} \Biggr] + \frac{\dot \omega_{k} \ddot \omega_{k}}{\omega_{k} ^5} \Biggl[ \left(-\frac{21 \xi }{4}+\frac{21}{16}\right) \frac{\dot{a}}{a} \notag\\
    &+ \left(-\frac{7 \xi }{4}+\frac{7}{16}\right) \frac{\dot{b }}{b } \Biggr] + \frac{\ddot \omega_{k}^2}{32 \omega_{k} ^5} + \frac{\omega_{k}^{(3)}}{\omega_{k} ^4} \Biggl[ \left(\frac{3 \xi }{4}-\frac{3}{16}\right) \frac{\dot{a}}{a} + \left(\frac{\xi }{4}-\frac{1}{16}\right) \frac{\dot{b }}{b } \Biggr] 
    -\frac{\dot \omega_{k} \omega_{k}^{(3)}}{16 \omega_{k} ^5}\label{eq:NCIntegrand4}
\end{align}
Dimensional analysis shows that the number of time derivatives in each term is equal to its adiabatic order, as expected. Plugging in the expression \eqref{eq:NComega} for $\omega_k$, the adiabatic expansions for the integrands become sums of terms involving a coefficient depending on geometric quantities ($a$, $b$, $\dot a$, $\dot b$,...) times an expression of the form
\begin{align}
    \frac{(k_4)^\kappa (k_T)^\lambda}{\omega_{k}^\delta}
    \label{eq:NCPreInt}
\end{align}
where $\kappa$ and $\lambda$ are non-negative even numbers, while $\delta$ is an odd number greater than or equal to $-1$. For example, the last term in \eqref{eq:NCIntegrand2} can be expressed as
\begin{align}
    \frac{\dot \omega^2}{\omega^3} = \frac{\dot a^2}{a^6} \frac{k_T^4}{\omega^5} + 
    \frac{2 \dot a \dot b}{a^3 b^3}\frac{k_T^2 \, k_4^2}{\omega^5} + 
    \frac{\dot b^2}{b^6}\frac{k_4^4}{\omega^5}
\end{align}
Similar adiabatic integrand expansions can be obtained for the other components. If $\delta$ is not large enough, then naively integrating \eqref{eq:NCPreInt} over the momenta results in divergences. Using dimensional regularization yields an expression in terms of gamma functions (see Appendix~\ref{sec:PNCAppendix} for details).
\begin{align}
    P_{\,\rm{NC}} ( d,\kappa,\lambda,\delta ) : \! &= \int_{-\infty}^{\infty} d^d k \, \frac{(k_d)^\kappa (k_T)^\lambda}{\left[ \left( \frac{k_T}{a} \right)^2 + \left( \frac{k_d}{b} \right)^2 + m^2 \right]^{\delta/2}}\\
    &\hspace{-10mm}= \pi^{(d-1)/2} \left( \frac{1+(-1)^\kappa}{2} \right) \frac{\Gamma \! \left( \frac{\kappa+1}{2}\right) \Gamma \! \left( -\frac{d+\kappa+\lambda-\delta}{2}\right) \Gamma \! \left( \frac{d+\lambda-1}{2}\right)}{\Gamma \! \left( \frac{d-1}{2}\right) \Gamma \! \left( \frac{\delta}{2}\right)} \ a^{d+\lambda-1} b^{\kappa+1} \, m^{d+\kappa+\lambda-\delta} \label{eq:PNC}
\end{align}
As such, summing over the adiabatic orders $l$, we can write 
\begin{align}
    \vev{T_{A B}}_{\rm{A}} &= \sum_{l=0}^{5} \vev{T_{A B}}_{\rm{A}}^{( l )} = \sum_{l=0}^{5} \sum_{\kappa,\lambda,\delta} A_{A B}^{( l )} ( 4, \kappa, \lambda, \delta ) \, \lim_{\epsilon \to 0} P_{\,\rm{NC}} ( 4 + \epsilon, \kappa, \lambda, \delta ) \label{eq:NCTalphabetaExpansion}
\end{align}
where the $A_{A B}^{( l )} ( d, \kappa, \lambda, \delta )$ coefficients depend only on geometric quantities.

Next, following similar steps as the non-compact case, the formal energy density \eqref{eq:genericTVEV} may be obtained by replacing the measure and relevant quantities in the adiabatic integrand expressions such as \eqref{eq:bianchiNCTVEV}.
\begin{align}
    \int_{-\infty}^{\infty} d^4 k \rightarrow \frac{1}{r_0} \sum_{n = -\infty}^{\infty} \int_{-\infty}^{\infty} d^3 k, \quad W_{k} \rightarrow W_{nk}, \quad \omega_{k} \rightarrow \omega_{nk}
\end{align}
This allows us to write
\begin{align}
    \vev{T_{A B}} = \sum_{l=0}^{5} \sum_{\kappa,\lambda,\delta} A_{A B}^{( l )} ( 4, \kappa, \lambda, \delta ) \, \lim_{\epsilon \to 0} P_{\rm{C}} ( 4 + \epsilon, \kappa, \lambda, \delta )
\end{align}
where the $A_{A B}^{( l )} ( d, \kappa, \lambda, \delta )$ are the same as in \eqref{eq:NCTalphabetaExpansion}, and the regularized sum/integral (see Appendix~\ref{sec:PCAppendix} for details) is
\begin{align}
    P_{\rm{C}} ( d, \kappa, \lambda, \delta ) : \! &= \frac{1}{r_0} \sum_{n = -\infty}^{\infty} \int_{-\infty}^{\infty} \, d^{d-1} k \, \frac{\left( n/r_0 \right) ^\kappa \left( k_T \right)^\lambda}{\left[ \left( \frac{k_T}{a} \right)^2 + \left( \frac{n/r_0}{b} \right)^2 + m^2 \right]^{\delta/2}}\\
    &= \pi^{\left( d - 1 \right) / 2} \ \frac{\Gamma \! \left( -\frac{d+\lambda-\delta-1}{2}\right) \Gamma \! \left( \frac{d+\lambda-1}{2}\right)}{\Gamma \! \left( \frac{d - 1}{2} \right) \Gamma \! \left( \frac{\delta}{2} \right)} a^{d+\lambda-1} \, r_{0}^{-d-\kappa-\lambda+\delta} \, b^{-d-\lambda+\delta+1} \, S ( d, \kappa, \lambda, \delta ) \label{eq:PC}\\
    S ( d, \kappa, \lambda, \delta ) : \! &= \sum_{p=0}^{\kappa/2} \left( -1 \right)^p \binom{\kappa/2}{p} \left( r_0 m b \right)^{2p} \zeta \! \left(s = p - \frac{\kappa}{2} - \frac{d+\lambda-\delta-1}{2}, \ \alpha = r_0 m b \right) \label{eq:combS}\\
    \zeta ( s, \alpha ) : \! &= \sum_{n=-\infty}^{\infty} \left( n^2 + \alpha^2 \right)^{-s}
\end{align}
The Epstein-Hurwitz zeta function $\zeta ( s, \alpha )$ is finite for $\mathfrak{R}(s) > \frac{1}{2}$ and divergent elsewhere. Zeta regularization yields the following analytic continuation \cite{Elizalde:1988rh}
\begin{align}
    \zeta ( s, \alpha ) = \frac{\pi^{1/2}}{\alpha^{2s-1} \, \Gamma ( s )} \left[ \Gamma \! \left( s - \frac{1}{2} \right) + 4 \sum_{n=1}^{\infty} \left( n \pi \alpha \right)^{s-\frac{1}{2}} K_{s-\frac{1}{2}} ( 2 n \pi \alpha ) \right] \label{eq:zetaAC}
\end{align}

The expression for $P_{\rm{C}} ( 4 + \epsilon, \kappa, \lambda, \delta )$ can be separated into a term that is independent of $r_0$, as well as other $r_0$-dependent terms that involve summations over Bessel functions. The $r_0$-independent term is precisely the same as $P_{\,\rm{NC}} ( 4 + \epsilon, \kappa, \lambda, \delta )$ and can be interpreted as the regularized UV divergence that is removed by the adiabatic subtraction procedure. As for the other $r_0$-dependent pieces involving the Bessel functions, the poles of the first gamma function in the numerator of \eqref{eq:PC} and the one in the prefactor of \eqref{eq:zetaAC} cancel in the $\epsilon \to 0$ limit. We can then define a regulated sum/integral which is of the form (neglecting numerical and geometric coefficients)
\begin{align}
    P_{\rm{reg}} ( 4, \kappa, \lambda, \delta ) : \! &= \lim_{\epsilon \to 0 } \left( P_{\rm{C}} ( 4 + \epsilon, \kappa, \lambda, \delta ) 
    - P_{\,\rm{NC}} ( 4 + \epsilon, \kappa, \lambda, \delta ) \right)\\
    &\sim m^{4+\kappa+\lambda-\delta} \sum_{p=0}^{\kappa/2} \, Q \! \left( s = p - \frac{\kappa}{2} - \frac{4+\lambda-\delta-1}{2}, \, \alpha = r_0 m b \right) \label{eq:PRegForm}
\end{align}
where
\begin{align}
    Q ( s, \alpha ) : \! &= \sum_{n=1}^{\infty} \left( n \pi \alpha \right)^{s-\frac{1}{2}} K_{s-\frac{1}{2}} ( 2 n \pi \alpha )
    \label{eq:QFunc}
\end{align}
Since the Bessel functions in \eqref{eq:QFunc} decay exponentially in $\alpha$, $P_{\rm{reg}} ( 4, \kappa, \lambda, \delta )$ decays exponentially in the mass $m$. They also decay exponentially in $n$, and hence, $P_{\rm{reg}} ( 4, \kappa, \lambda, \delta )$ is finite and well-defined for non-zero mass.

While the computation we have done so far is for the $d=4$ scenario, an analogous computation can be performed for $d=3$. From \eqref{eq:regVEVgeneric}, the regularized stress-energy expectation for generic $m > 0$ and $\xi$ may then be expressed as
\begin{align}
    \vev{T_{A B}}_{\rm{reg}} &= \sum_{l=0}^{d+1} \sum_{\kappa,\lambda,\delta} A_{A B}^{( l )} ( d, \kappa, \lambda, \delta ) \, P_{\rm{reg}} ( d, \kappa, \lambda, \delta ) \label{eq:TalphabetaVEVreg}
\end{align}
The geometric coefficients $A_{A B}^{( l )} ( d, \kappa, \lambda, \delta )$ are too long and numerous to list in this article.\footnote{They can be found in Mathematica notebooks at \url{https://github.com/apaul1999/Time-Dependent-Dilaton}.} A table containing the values of $P_{\rm{reg}} ( d, \kappa, \lambda, \delta )$ for all relevant indices can be found in Appendix~\ref{sec:PRegValsappendix} for both $d=3$ and $d=4$. With these expressions, one can check that the conservation equation is satisfied at each adiabatic order.
\begin{align}
    J^B = \nabla_A \vev{T^{A B}}_{\rm{reg}} = 0
    \label{eq:conservationeq}
\end{align}

\section{The massless, conformal limit of \texorpdfstring{$\bm{\vev{T_{A B}}_{\rm{reg}}}$}{}} \label{sec:MasslessConformalLimit}
Now, we study the massless, conformally coupled limit ($m \to 0, \, \xi \to \frac{d-1}{4 d}$) as we would like to check the results we have obtained using our modified prescription against standard results in the literature, most of which are reported in this specific limit. However, we must be careful as additional divergences arise in the massless case, in contrast to the finite results we have obtained for the massive case. Fermions might cancel some of these divergences if they were included in our computation, as in \cite{Katayama:2026ras}. However, these cancellations are not necessary, as we show in this section and in section~\ref{sec:EinsteinFrame}, because the divergences can be readily absorbed into counterterms. 

\subsection{\texorpdfstring{$d = 3$}{} scenario} \label{sec:d3Scenario}
Let us set $\xi = \frac{1}{6}$. While expressions of the form \eqref{eq:PRegForm} are finite for non-zero mass, we encounter additional logarithmic divergences in the massless limit. Furthermore, for $d=3$, it is not possible to explicitly carry out the sum in \eqref{eq:QFunc} prior to taking $m \to 0$. Hence, we perform a resummation by interchanging the order of the massless limit and the mode summation. The terms at $0$-th and $2$-nd adiabatic order have the same finite value regardless of the order of operations and are well-defined. However, at $4$-th order, the summand consists of a part that is finite and independent of $n$ in the massless limit (referred to as $S^{\rm{fin}}_{A B}$), as well as a piece that still diverges logarithmically in $m$ (referred to as $S^{\rm{div}}_{A B}$). This latter divergent summand may be expressed as a linear combination $S^{\rm{div}}_{A B} = c_1 \prescript{(1)}{}{\mathcal{H}}_{A B} + c_2 \prescript{(2)}{}{\mathcal{H}}_{A B}$, where
\begin{align}
    \prescript{(1)}{}{\mathcal{H}}_{A B} : \! &= \frac{1}{|g|^{1/2}} \frac{\delta}{\delta g^{A B}} \int |g|^{1/2} \,\mathcal{R}^2 \,d^{4} x \notag\\
    &= 2 \, \nabla_B \nabla_A \mathcal{R} - 2 \, g_{A B} \,\Box \, \mathcal{R} - \frac{1}{2} g_{A B} \mathcal{R}^2 + 2 \, \mathcal{R} \, \mathcal{R}_{A B}\\
    \prescript{(2)}{}{\mathcal{H}}_{A B} : \! &= \frac{1}{|g|^{1/2}} \frac{\delta}{\delta g^{A B}} \int |g|^{1/2} \,\mathcal{R}^{M N} \mathcal{R}_{M N} \,d^{4} x \notag\\
    &= \nabla_B \nabla_A \mathcal{R} - \frac{1}{2} g_{A B} \Box \, \mathcal{R} - \Box \, \mathcal{R}_{A B} - \frac{1}{2} g_{A B} \mathcal{R}_{C D} \mathcal{R}^{C D} + 2 \, \mathcal{R}_{C A D B} \mathcal{R}^{C D}
\end{align}
and the coefficients are given by 
\begin{align}
    c_1 = -\frac{K_{0} \left( 2 n \pi r_0 m b \right)}{720 \pi^2}, \, c_2 = \frac{K_{0} \left( 2 n \pi r_0 m b \right)}{240 \pi^2}
\end{align}
The Bessel function can be expanded in the small mass limit as
\begin{align}
    K_{0} \left( 2 n \pi r_0 m b \right) = -\left[\gamma + \log(b)\right] - \log(n \pi r_0 m) + \mathcal{O}(m^1)
\end{align}
Thus, the piece of $S^{\rm{div}}_{A B}$ that is proportional to $\log(n \pi r_0 m)$ may be discarded since it can be interpreted as renormalizing the couplings of higher curvature terms\footnote{A term in the gravitational action of the form $\mathcal{R}^{M N P Q} \mathcal{R}_{M N P Q}$ is not necessary for renormalization in $(1+3)$-dimensional spacetime, since by the Gauss-Bonnet theorem \cite{Chern:1945} we have
\begin{align}
    \mathcal{H}_{A B} : \! &= \frac{1}{|g|^{1/2}} \frac{\delta}{\delta g^{A B}} \int |g|^{1/2} \,\mathcal{R}^{M N P Q} \mathcal{R}_{M N P Q} \,d^{4} x = -\prescript{(1)}{}{\mathcal{H}}_{A B} + 4 \prescript{(2)}{}{\mathcal{H}}_{A B} \notag
\end{align}} that have been added to the gravitational action \cite{Birrell:1982ix} 
\begin{align}
    S = \int d^{4} x \, |g|^{1/2} \left[ c_1 \mathcal{R}^2 + c_2 \mathcal{R}^{M N} \mathcal{R}_{M N} \right]
\end{align}
As for the piece of $S^{\rm{div}}_{A B}$ proportional to $-\left[\gamma + \log(b)\right]$, it is finite and independent of $n$ in the massless limit, and hence, must also be added to $S^{\rm{fin}}_{A B}$. Carrying out the mode summation over this total summand that is constant in $n$ requires zeta regularization, which is achieved by multiplying the summand with $\zeta(0) = -\frac{1}{2}$. This gives us 
\begin{align}
    \vev{T_{A B}}_{\rm{reg}}^{\left( 4 \right)} = -\frac{1}{2} S^{\rm{fin}}_{A B} + \frac{\gamma + \log(b)}{480 \pi^2} \left( -\frac{1}{3} \prescript{(1)}{}{\mathcal{H}}_{A B} + \prescript{(2)}{}{\mathcal{H}}_{A B} \right)
\end{align}
Note that there is a degree of arbitrariness due to the scheme dependence of the renormalization procedure described above. With this scheme choice, the explicit expressions for the regularized stress-energy tensor in the massless, conformally coupled limit, organized by adiabatic order, are as follows.
\begin{align}
    \vev{T_{00}}_{\rm{reg}}^{\left( 0 \right)} =& -\frac{1}{1440 \, \pi^2 \, r_0^4 \, b^4}\\
    \vev{T_{00}}_{\rm{reg}}^{\left( 2 \right)} =& \, 0\\
    \vev{T_{00}}_{\rm{reg}}^{\left( 4 \right)} =& \, \frac{1}{2880 \, \pi^2} \biggl( -\frac{8 \dot{a}^3 \dot{b }}{a^3 b }+\frac{8 \dot{a}^2 \dot{b }^2}{a^2 b ^2}-\frac{4 \dot{a} \dot{b } \ddot{a}}{a^2 b}+\frac{8 \dot{a}^2 \ddot{b }}{a^2 b }-\frac{8 \dot{a} \dot{b }^3}{a b ^3}+\frac{8 \dot{b }^2 \ddot{a}}{a b ^2}-\frac{4 \dot{a} \dot{b } \ddot{b }}{a b^2}-\frac{4 \ddot{a} \ddot{b }}{a b } \notag\\
    &+\frac{4 \dot{b } a^{(3)}}{a b } +\frac{4 \dot{a} b^{(3)}}{a b } +\frac{2 \dot{b }^4}{b^4}-\frac{2 \dot{b }^2 \ddot{b }}{b ^3}+\frac{\ddot{b }^2}{b ^2}-\frac{2 \dot{b } b^{(3)}}{b ^2} \biggr) + \frac{\gamma+\log(b)}{1440 \pi^2} \biggl( \frac{\dot{a}^4}{a^4}-\frac{4 \dot{a}^3 \dot{b}}{a^3 b} \notag\\
    &+\frac{5 \dot{a}^2 \dot{b}^2}{a^2 b^2} +\frac{4 \dot{a}^2 \ddot{b}}{a^2 b} -\frac{2 \dot{a} \dot{b} \ddot{a}}{a^2 b} +\frac{\ddot{a}^2}{a^2} -\frac{2 \dot{a} a^{(3)}}{a^2}-\frac{2 \dot{a} \dot{b}^3}{a b^3}-\frac{6 \dot{a} \dot{b} \ddot{b}}{a b^2}+\frac{2 \dot{b}^2 \ddot{a}}{a b^2}-\frac{2 \ddot{a} \ddot{b}}{a b} \notag\\
    &+\frac{2 \dot{a} b^{(3)}}{a b} +\frac{2 \dot{b} a^{(3)}}{a b} +\frac{2 \dot{b}^2 \ddot{b}}{b^3} +\frac{\ddot{b}^2}{b^2} -\frac{2 \dot{b} b^{(3)}}{b^2} \biggr)\\
    \vev{T_{ii}}_{\rm{reg}}^{\left( 0 \right)} =& \, \frac{a^2}{1440 \, \pi^2 \, r_0^4 \, b^4}\\
    \vev{T_{ii}}_{\rm{reg}}^{\left( 2 \right)} =& \, 0\\
    \vev{T_{ii}}_{\rm{reg}}^{\left( 4 \right)} =& \, \frac{1}{2880 \, \pi^2} \biggl( -\frac{2 a^2 \dot{b }^4}{b ^4}-\frac{2 a^2 b ^{(4)}}{b}+\frac{6 a^2 \dot{b }^2 \ddot{b }}{b ^3}+\frac{a^2 \ddot{b }^2}{b ^2}-\frac{2 a^2 \dot{b } b^{(3)}}{b ^2}-\frac{10 a \dot{a} \dot{b} \ddot{b }}{b ^2} \notag\\
    &+\frac{4 \dot{a}^2 \dot{b }^2}{b ^2} +\frac{8 \dot{a} \dot{b } \ddot{a}}{b } +\frac{4 \dot{a}^2 \ddot{b }}{b }-\frac{4 a \ddot{a} \ddot{b }}{b }+\frac{4 a \dot{b } a^{(3)}}{b }-\frac{6 a \dot{a} b^{(3)}}{b }-\frac{4 \dot{a}^3 \dot{b }}{a b } \biggr) \notag\\
    &+ \frac{\gamma+\log(b)}{1440 \pi^2} \biggl( a a^{(4)} -\frac{a^2 b^{(4)}}{b}-\frac{a^2 \dot{b}^2 \ddot{b}}{b^3} +\frac{2 a^2 \ddot{b}^2}{b^2}+\frac{a^2 \dot{b} b^{(3)}}{b^2}+\frac{\dot{a}^4}{a^2}+\frac{a \dot{a} \dot{b}^3}{b^3} \notag\\
    &-\frac{2 a \dot{a} \dot{b} \ddot{b}}{b^2} -\frac{a \dot{b}^2 \ddot{a}}{b^2}+\frac{2 \dot{a}^2 \ddot{b}}{b} +\frac{4 \dot{a} \dot{b} \ddot{a}}{b}-\frac{2 a \ddot{a} \ddot{b}}{b} -\frac{3 a \dot{a} b^{(3)}}{b} +\frac{2 a \dot{b} a^{(3)}}{b}-\frac{2 \dot{a}^3 \dot{b}}{a b} \notag\\
    &-\frac{2 \dot{a}^2 \ddot{a}}{a} \biggr)\\
    \vev{T_{33}}_{\rm{reg}}^{\left( 0 \right)} =& -\frac{1}{1440 \, \pi^2 \, r_0^4 \, b^2}\\
    \vev{T_{33}}_{\rm{reg}}^{\left( 2 \right)} =& \, 0\\
    \vev{T_{33}}_{\rm{reg}}^{\left( 4 \right)} =& \, \frac{1}{1440 \, \pi^2} \biggl( -\frac{\dot{a}^4 b ^2}{a^4}-\frac{2 a^{(4)} b ^2}{a}+\frac{4 \dot{a}^2 b ^2 \ddot{a}}{a^3}+\frac{b ^2 \ddot{a}^2}{a^2}-\frac{2 \dot{a} b^2 a^{(3)}}{a^2}-\frac{10 \dot{a} b  \dot{b } \ddot{a}}{a^2} \notag\\
    &+\frac{3 \dot{a}^2 \dot{b }^2}{a^2} +\frac{12 \dot{a} \dot{b } \ddot{b }}{a} +\frac{6 \dot{b }^2 \ddot{a}}{a}-\frac{4 b  \ddot{a} \ddot{b }}{a}+\frac{4 \dot{a} b  b^{(3)}}{a}-\frac{6 b  \dot{b } a^{(3)}}{a}-\frac{6 \dot{a} \dot{b }^3}{a b }+b  b ^{(4)} \notag\\
    &+\frac{3 \dot{b }^4}{b^2} +\frac{3 \ddot{b }^2}{2} -\frac{8 \dot{b }^2 \ddot{b }}{b }+2 \dot{b } b^{(3)} \biggr) + \frac{\gamma+\log(b)}{1440 \pi^2} \biggl( -\frac{\dot{a}^4 b^2}{a^4}-\frac{2 a^{(4)} b^2}{a}+\frac{4 \dot{a}^2 b^2 \ddot{a}}{a^3} \notag\\
    &+\frac{b^2 \ddot{a}^2}{a^2} -\frac{2 \dot{a} b^2 a^{(3)}}{a^2}-\frac{10 \dot{a} b \dot{b} \ddot{a}}{a^2} +\frac{5 \dot{a}^2 \dot{b}^2}{a^2} -\frac{2 \dot{a} \dot{b} \ddot{b}}{a}+\frac{4 \dot{b}^2 \ddot{a}}{a}+\frac{2 b \ddot{a} \ddot{b}}{a}+\frac{8 \dot{a} b b^{(3)}}{a} \notag\\
    &-\frac{2 b \dot{b} a^{(3)}}{a} -\frac{4 \dot{a} \dot{b}^3}{a b}+2 b b^{(4)} -3 \ddot{b}^2+\frac{4 \dot{b}^2 \ddot{b}}{b} -4 \dot{b} b^{(3)} \biggr)
\end{align}

The $0$-th order expressions, which may be interpreted as the Casimir energy/pressures, match known results for the time-independent case \cite{Arkani-Hamed:2007ryu, Parker:2009uva}. The conservation equation \eqref{eq:conservationeq} is satisfied at each adiabatic order. In the limit where $a(t) = b(t)$, the stress-energy tensors at each order match those obtained in the conformally flat scenario using the standard adiabatic regularization prescription (Appendix~\ref{sec:AdExactVals3dappendix}). Notice that the renormalization scheme dependent terms vanish in this limit. Finally, computing the trace (whose entire contribution comes from $4$-th order terms), one obtains
\begin{align}
    \vev{{T^{A}}_{A}}_{\rm{reg}} =& \, \vev{{T^{A}}_{A}}^{\left(4\right)}_{\rm{reg}} = \, \frac{1}{1440 \pi^2} \biggl( \frac{\dot{a}^4}{a^4}+\frac{2 a^{(4)}}{a}-\frac{4 \dot{a}^2 \ddot{a}}{a^3}-\frac{3 \dot{a}^2 \dot{b }^2}{a^2 b ^2}-\frac{\ddot{a}^2}{a^2}+\frac{2 \dot{a} a^{(3)}}{a^2}+\frac{2 \dot{a} \dot{b }^3}{a b ^3} \notag\\
    &-\frac{4 \dot{a} \dot{b } \ddot{b }}{a b^2} -\frac{2 \dot{b }^2 \ddot{a}}{a b^2} +\frac{6 \ddot{a} \ddot{b }}{a b } +\frac{4 \dot{a} b^{(3)}}{a b }+\frac{4 \dot{b } a^{(3)}}{a b }+\frac{b ^{(4)}}{b}+\frac{\dot{b }^2 \ddot{b }}{b ^3}-\frac{2 \ddot{b }^2}{b ^2}-\frac{\dot{b } b^{(3)}}{b ^2} \biggr) \label{eq:d4tracemasslessconformal}
\end{align}
It is clear that the scheme dependent terms cancel each other in the trace. One can check that \eqref{eq:d4tracemasslessconformal} reproduces the well known trace anomaly in ($1+3$)-dimensional FLRW spacetime \cite{Wald:1978pj, Birrell:1982ix, Parker:2009uva} upon taking $b(t) \to a(t)$.

\subsection{\texorpdfstring{$d = 4$}{} scenario} \label{sec:d4Scenario}

After setting $\xi = \frac{3}{16}$, we use the resummation procedure (taking $m \to 0$ prior to carrying out the mode summation over $n$) again. As before, the resummation leaves the $0$-th and $2$-nd adiabatic order contributions to the stress-energy tensor invariant and finite.
\begin{align}
    \vev{T_{00}}_{\rm{reg}}^{\left( 0 \right)} =& -\frac{3 \, \zeta (5)}{128 \, \pi^7 \, r_0^5 \, b^5}\\
    \vev{T_{00}}_{\rm{reg}}^{\left( 2 \right)} =& \, \frac{\zeta(3)}{512 \, \pi^5 \, r_{0}^3 b^3} \biggl( \frac{\dot a^2}{a^2} - \frac{2 \dot a \dot b}{a b} + \frac{\dot b^2}{b^2} \biggr)\\
    \vev{T_{ii}}_{\rm{reg}}^{\left( 0 \right)} =& \, \frac{3 \, \zeta (5) a^2}{128 \, \pi^7 \, r_0^5 \, b^5}\\
    \vev{T_{ii}}_{\rm{reg}}^{\left( 2 \right)} =& \, \frac{\zeta(3)}{1536 \, \pi^5 \, r_{0}^3 b^3} \biggl( -\frac{3 a^2 \dot{b }^2}{b ^2}+\frac{2 a^2 \ddot{b }}{b }+\frac{4 \dot{a} a \dot{b }}{b }-2 a \ddot{a}-\dot{a}^2 \biggr)\\
    \vev{T_{44}}_{\rm{reg}}^{\left( 0 \right)} =& -\frac{3 \, \zeta (5)}{32 \, \pi^7 \, r_0^5 \, b^3}\\
    \vev{T_{44}}_{\rm{reg}}^{\left( 2 \right)} =& \, \frac{\zeta(3)}{256 \, \pi^5 \, r_{0}^3 b^3} \biggl( \frac{\dot{a}^2 b ^2}{a^2}+\frac{b ^2 \ddot{a}}{a}-\frac{3 \dot{a} \dot{b } b }{a}-b  \ddot{b }+2 \dot{b}^2 \biggr)
\end{align}

At $4$-th adiabatic order, taking the massless limit of the summand first results in an expression that is proportional to $n^{-1}$. Upon expressing the $N$-th partial sum as $S_N =\sum_{n=1}^N n^{-1} = \gamma + \log(N) + \epsilon_N$ (where $\lim_{N \to \infty} \epsilon_N \to 0$), we can see that there is still a logarithmic divergence when $N \to \infty$. Unlike the $d=3$ scenario, further zeta regularization\footnote{The divergent sum $S_\infty =\sum_{n=1}^\infty n^{-1}$ cannot be zeta regularized as it corresponds to evaluating the Riemann zeta function $\zeta(z)$ at $z=1$, a simple pole.} or direct renormalization by the addition of suitable higher curvature counterterms to the $(1+4)$-dimensional Jordan frame gravitational action cannot be used to deal with this divergence. However, as we will see in section~\ref{sec:EinsteinFrame}, these divergences can still be renormalized by operators in the dimensionally-reduced Einstein frame action. With our specific scheme choice of keeping the terms proportional to $\gamma$, the $4$-th adiabatic order regularized results are as follows.
\begin{align}
    \vev{T_{00}}_{\rm{reg}}^{\left( 4 \right)} =& \, \frac{\gamma}{4096 \, \pi^3 \, r_{0} b} \biggl( \frac{3 \dot{a}^4}{a^4}-\frac{10 \dot{a}^3 \dot{b }}{a^3 b }-\frac{2 \dot{a}^2 \ddot{a}}{a^3}+\frac{11 \dot{a}^2 \dot{b }^2}{a^2 b ^2}+\frac{6 \dot{a}^2 \ddot{b }}{a^2 b }+\frac{2 \dot{a} \dot{b } \ddot{a}}{a^2 b }+\frac{\ddot{a}^2}{a^2}-\frac{2 \dot{a} a^{(3)}}{a^2} \notag\\
    &-\frac{4 \dot{a} \dot{b }^3}{a b ^3} -\frac{10 \dot{a} \dot{b } \ddot{b }}{a b ^2}-\frac{2 \ddot{a} \ddot{b }}{a b }+\frac{2 \dot{a} b^{(3)}}{a b}+\frac{2 \dot{b } a^{(3)}}{a b }+\frac{4 \dot{b }^2 \ddot{b }}{b^3}+\frac{\ddot{b }^2}{b ^2}-\frac{2 \dot{b } b^{(3)}}{b^2} \biggr)\\
    \vev{T_{ii}}_{\rm{reg}}^{\left( 4 \right)} =& \, \frac{\gamma}{6144 \, \pi^3 \, r_{0} b} \biggl( a a^{(4)}-\frac{a^2 b ^{(4)}}{b }-\frac{4 a^2 \dot{b }^2 \ddot{b }}{b ^3}+\frac{7 a^2 \ddot{b }^2}{2 b ^2}+\frac{3 a^2 \dot{b } b^{(3)}}{b ^2}+\frac{3 \dot{a}^4}{2 a^2}+\frac{4 a \dot{a} \dot{b }^3}{b ^3} \notag\\
    &+\frac{a \dot{a} \dot{b } \ddot{b }}{b ^2} -\frac{2 a \dot{b }^2 \ddot{a}}{b ^2} -\frac{11 \dot{a}^2 \dot{b }^2}{2 b ^2}+\frac{2 \dot{a}^2 \ddot{b }}{b}+\frac{9 \dot{a} \dot{b } \ddot{a}}{b}-\frac{5 a \ddot{a} \ddot{b }}{b }-\frac{5 a \dot{a} b^{(3)}}{b }-\frac{6 \dot{a}^2 \ddot{a}}{a} \notag\\
    &+\frac{3 \ddot{a}^2}{2}+2 \dot{a} a^{(3)} \biggr) \\
    \vev{T_{44}}_{\rm{reg}}^{\left( 4 \right)} =& \, \frac{\gamma}{2048 \, \pi^3 \, r_{0} b} \biggl( -\frac{a^{(4)} b ^2}{a}+\frac{5 \dot{a}^2 b ^2 \ddot{a}}{a^3}-\frac{5 \dot{a}^3 b  \dot{b }}{a^3}-\frac{b ^2 \ddot{a}^2}{a^2}-\frac{3 \dot{a} b^2 a^{(3)}}{a^2}+\frac{\dot{a}^2 b  \ddot{b }}{a^2}-\frac{8 \dot{a} b  \dot{b } \ddot{a}}{a^2} \notag\\
    &+\frac{11 \dot{a}^2 \dot{b }^2}{a^2} -\frac{6 \dot{a} \dot{b } \ddot{b }}{a} +\frac{2 \dot{b }^2 \ddot{a}}{a}+\frac{4 b  \ddot{a} \ddot{b }}{a}+\frac{6 \dot{a} b  b^{(3)}}{a}+\frac{b  \dot{b } a^{(3)}}{a}-\frac{6 \dot{a} \dot{b }^3}{a b }+b b ^{(4)}-3 \ddot{b }^2 \notag\\
    &+\frac{6 \dot{b }^2 \ddot{b }}{b}-4 \dot{b } b^{(3)} \biggr)
\end{align}

Similar to the $d=3$ case, the $0$-th order Casimir terms match known time-independent results \cite{Appelquist:1982zs, Appelquist:1983vs, Appelquist:1983wx, Weinberg:1983xy, Candelas:1983ae, Kantowski:1986iv}. The stress-energy tensor satisfies the conservation equation \eqref{eq:conservationeq} at each adiabatic order. It is also traceless at each order,
\begin{align}
    \vev{{T^{A}}_{A}}_{\rm{reg}} = 0
    \label{eq:tracelesseq}
\end{align}
as is expected for a massless, conformally coupled field in a spacetime with an odd number of dimensions \cite{Birrell:1982ix, Parker:2009uva}. 
Finally, in the limit $a(t) = b(t)$, the stress-energy expressions at each order match those obtained in the conformally flat scenario using the standard adiabatic regularization prescription (Appendix~\ref{sec:AdExactVals4dappendix}). The $2$-nd and $4$-th order contributions vanish in this limit. This agreement with known results suggests that our modified prescription is likely valid.

\section{The dimensionally-reduced Einstein frame action for \texorpdfstring{$\bm{d = 4}$}{}} \label{sec:EinsteinFrame}
The computations we have presented so far for $d=4$ have all been performed in the Jordan frame with the higher dimensional action \eqref{eq:d_action}. However, it is interesting to consider the dynamics of the system from the physically relevant $(1+3)$-dimensional point of view. Furthermore, as mentioned in section~\ref{sec:d4Scenario}, the divergences that we encountered in the massless and conformally-coupled limit may be renormalized by terms within the dimensionally-reduced action in the Einstein frame. Directly computing this action is difficult, so in order to reconstruct it, we will work at the level of the gravitational field equations.

First, we must identify the correct transformations to the Einstein frame. The metric \eqref{eq:d_FLRW} for $d=4$ can be decomposed as
\begin{align}
    ds^2 &= g_{A B} \, dx^A \,dx^B = ds_4^2 - b(t)^2 dx_{4}^2
\end{align}
where
\begin{align}
    ds_4^2 &= \bar{g}_{\mu \nu} dx^\mu dx^\nu = dt^2 - a(t)^2 \sum_{j=1}^{3} dx_j^2, \hspace{5mm} \mu,\nu \in \left\{ 0, ..., 3 \right\}
\end{align}
Consider the following change of variables,
\begin{align}
    \tilde{g}_{\mu \nu} &= b(t) \bar{g}_{\mu \nu} \label{eq:gtilde}\\
    \tilde{t} &= \int_{0}^{t} b^{1/2}(t') \, dt' \label{eq:ttilde}\\
    \tilde{a}(\tilde{t}) &= a(t), \quad \tilde{b}(\tilde{t}) = b(t), \quad \tilde{c}(\tilde{t}) = \tilde{b}^{1/2}(\tilde{t}) \, \tilde{a}(\tilde{t})\\
    M_4^2 &= (2 \pi r_0) M_5^3\\
    \tilde{\chi}(\tilde{t}) &= \sqrt{\frac{3}{2}} \, M_4 \ln\left(\tilde{b} \left( \tilde{t} \, \right)\right) \label{eq:chitilde}
\end{align}
where $M_4$ is the $(1+3)$-dimensional reduced Planck mass. The new line element can be expressed as
\begin{align}
    d\tilde{s}_4^2 &= \tilde{g}_{\tilde{\mu} \tilde{\nu}} dx^{\tilde{\mu}} dx^{\tilde{\nu}} = d \tilde{t}\,^2 - \tilde{c}(\tilde{t})^2 \sum_{j=1}^{3} dx_j^2
\end{align}
As shown in \cite{Carroll:2004st}, the Einstein-Hilbert term in the action \eqref{eq:d_action}, upon transforming to the Einstein frame, is
\begin{align}
    S = \int d^{4} \tilde{x} \, |\tilde{g}|^{1/2} \left[ \frac{M_4^2}{2} \tilde{\mathcal{R}}_{(4)} + \frac{1}{2} \tilde{g}_{\tilde{\mu} \tilde{\nu}} \tilde{\nabla}_{\tilde{\mu}} \tilde{\chi} \tilde{\nabla}_{\tilde{\nu}} \tilde{\chi}\right] \label{eq:Carrolldimred}
\end{align}
where $\tilde{\mathcal{R}}_{(4)} = \mathcal{R}\left[\tilde{g}_{\tilde{\mu} \tilde{\nu}}\right]$. The gravitational field equations in the Jordan frame are given by
\begin{align}
    \mathcal{R}_{(5) A B} - \frac{1}{2} \mathcal{R}_{(5)} \, g_{A B} = -{M_5^{-3}} \, T_{A B} \label{eq:5DJordanEinEq}
\end{align}
Plugging in the transformations \eqref{eq:gtilde}-\eqref{eq:chitilde}, dividing by common factors, and dropping all tildes for notational simplicity, we obtain three equations\footnote{Primes are used to denote derivatives with respect to $\tilde{t}$, as defined in \eqref{eq:ttilde}. However, we have dropped all tildes on the fields and coordinates for notational simplicity.} (the right hand sides are listed in Appendix~\ref{sec:RHSEFE})
\begin{align}
    3 \, \frac{\left(c'\right)^2}{c^2} -\frac{1}{2} \frac{\left(\chi'\right)^2}{M_4^2} &= T_1^{(0)}+T_1^{(2)}+T_1^{(4)} = T_1\label{eq:einframeeq1}\\
    -\frac{\left(c'\right)^2}{c^2}-\frac{1}{2} \frac{\left(\chi'\right)^2}{M_4^2} - 2 \, \frac{c''}{c} &= T_2^{(0)}+T_2^{(2)}+T_2^{(4)}  = T_2\label{eq:einframeeq2}\\
    -3 \, \frac{\left(c'\right)^2}{c^2} + 3\sqrt{\frac{3}{2}} \frac{c'}{c} \frac{\chi'}{M_4} -\frac{1}{2} \frac{\left(\chi'\right)^2}{M_4^2} - 3 \, \frac{c''}{c} + \sqrt{\frac{3}{2}} \frac{\chi''}{M_4} &= T_3^{(0)}+T_3^{(2)}+T_3^{(4)} = T_3\label{eq:einframeeq3}
\end{align}
Combining these equations yields
\begin{align}
    \chi'' + 3 \, \frac{c'}{c} \, \chi' - \sqrt{\frac{2}{3}} M_4 \left( \frac{1}{2} T_1 -\frac{3}{2} T_2 + T_3 \right) = 0 \label{eq:chicombo}
\end{align}

From \eqref{eq:Carrolldimred} and the structure of the terms in \eqref{eq:einframeeq1}-\eqref{eq:chicombo}, we can construct the part of the dimensionally reduced Einstein frame action that reproduces the gravitational field equations. Defining $X :\,= \frac{1}{2} g^{\mu\nu} \nabla_\mu\chi \nabla_\nu\chi \,$, we get
\begin{align}
    S_{\rm{E}} \supseteq& \int d^{4} x \, |g|^{1/2} \Bigl[ \frac{1}{2} f(\chi) g^{\mu \nu} \nabla_\mu \chi \, \nabla_\nu \chi -V\left( \chi \right) + \frac{M_{4}^{2}}{2} \mathcal{R}_{(4)} \left(h(\chi) + 1\right) + c_1 \mathcal{R}_{(4)}^2 + c_2 \left( \Box \chi \right)^2 \notag\\
    &\hspace{21mm}+ c_3 X^2 + c_4 X \Box \chi + c_5 \chi \Box \mathcal{R}_{(4)} + c_6 X \mathcal{R}_{(4)} \Bigr] \label{eq:EFaction}
\end{align}
Computing the Euler-Lagrange equation for $\chi(t)$ and comparing it to \eqref{eq:chicombo}, we identify
\begin{align}
    f\left(\chi\right) &= 1 + \frac{3\,\zeta(3)}{256 \, \pi^4 \, r_0^2 \, M_4^2} e^{-\sqrt{6} \chi / M_4}\\
    h\left(\chi\right) &= -\frac{\zeta(3)}{768 \, \pi^4 \, r_0^2 \, M_4^2} e^{-\sqrt{6} \chi / M_4}\\
    V\left(\chi\right) &= -\frac{3\,\zeta(5)}{64 \, \pi^6 \, r_0^4} e^{-2 \sqrt{6} \chi / M_4}
\end{align}
and fix the coupling constants
\begin{align}
    c_2 = \frac{3 \gamma}{4096 \pi^2 M_4^2} \, , \quad c_3 = \frac{9 \gamma}{2048 \pi^2 M_4^2} \, , \quad c_4 = -\frac{3 \sqrt{\frac{3}{2}}\gamma}{1024 \pi^2 M_4^3} \notag
\end{align}
\begin{align}
    c_5 = -\frac{\gamma}{2048 \sqrt{6} \pi^2 M_4} \, , \quad c_6 = \frac{\gamma}{2048 \pi^2 M_4^2}
\end{align}
While the action \eqref{eq:EFaction} reproduces the correct equation of motion for $\chi(t)$, we still need to check the equation of motion for $c(t)$. Computing the corresponding Euler-Lagrange equation, we find that it is a multiple of \eqref{eq:einframeeq2}, provided that we fix the constant
\begin{align}
    c_1 = \frac{\gamma}{73728 \pi^2}
\end{align}

Thus, we are able to recover the original Einstein field equations \eqref{eq:5DJordanEinEq} from the action above with the specified coupling constants. Furthermore, all the couplings that we have computed are proportional to $\gamma$. As we saw in section~\ref{sec:d4Scenario}, the terms proportional to $\gamma$ in the equations of motion are the finite portions of the formally divergent terms that we chose to keep through our renormalization scheme. Therefore, the remaining divergent pieces may be absorbed by the higher-order counterterms in \eqref{eq:EFaction}.

\section{Conclusion} \label{sec:Conclusion}
We have computed the vacuum expectation value of the stress-energy tensor of a scalar field in a spacetime that is the product of an FLRW background and a compact dimension ($\mathcal{M}^{1, \,d-1} \times \mathcal{S}^1$), where the size of the latter is allowed to vary with time. This computation has been performed for $d=3$ and $d=4$. To obtain a regularized expression for $\vev{T_{A B}}$, we propose a modification of the standard adiabatic regularization prescription that uses a WKB approximation for the field modes of the compact space rather than the exact solutions. We have checked that the expressions for $\vev{T_{A B}}_{\rm{reg}}$ are finite and covariantly conserved for generic non-zero mass and Ricci coupling. Additionally, we have found that in the massless and conformally coupled limit, our modified procedure requires further renormalization to obtain finite, covariantly conserved results. The computed expressions in this limit, when the FLRW and compact dimension scale factors are equal, reproduce known results for ($1+d$)-dimensional FLRW spacetime. Although we have not directly checked the equivalence of our method with point splitting in general, we have verified that they agree when the background spacetime is conformally flat. This is because our results agree with those obtained using the standard adiabatic regularization prescription, which has been proven to be equivalent to point splitting in the general case.

Having tested the method, the next step is to determine when the corrections become significant. In $1+4$ dimensions, the structure of the first non-adiabatic correction to the stress-energy tensor differs from the Einstein tensor $G_{A B}$, but is suppressed by powers of $1/{r_0 M_{5} b}$. Corrections are only expected to influence the time evolution when this factor approaches order one.

\acknowledgments

We have benefited from discussions with J. Distler, E. Elizalde, M. Montero, C. Vafa, and E. Verdaguer. This work was inspired by an earlier collaboration among SP, E. McDonough, D. Kaiser, and S. Sethi on a closely related project. SP has benefited from their insights. Part of the work of AP was supported by the Weinberg Institute for Theoretical Physics at the University of Texas at Austin. Part of the work of SP was performed at the Aspen Center for Physics, which is supported by National Science Foundation grant PHY-2210452.

\begin{appendices}

\section{Regularization of the \texorpdfstring{$\bm{P(d, \kappa, \lambda, \delta)}$}{} integrals} \label{sec:PAppendix}

\subsection{Non-compact} \label{sec:PNCAppendix}
\begin{align}
    P_{\,\rm{NC}} ( d,\kappa,\lambda,\delta ) : \! &= \int_{-\infty}^{\infty} dk_d \int_{-\infty}^{\infty} \, \prod_{i=1}^{d-1} dk_i \, \frac{(k_d)^\kappa (k_T)^\lambda}{\left[ \left( \frac{k_T}{a} \right)^2 + \left( \frac{k_d}{b} \right)^2 + m^2 \right]^{\delta/2}} \notag\\
    &\tag*{(substitute $\beta^2 = \left( \frac{k_d}{b} \right) ^2 + m^2)$} \notag\\
    &\hspace{-70pt}= \int_{-\infty}^{\infty} dk_d \ \left( k_d \right)^\kappa \, \frac{2 \pi^{\left( d - 1 \right) / 2}}{\Gamma \! \left( \frac{d - 1}{2} \right)} \int_{0}^{\infty} dk_T \, \frac{(k_T)^{d + \lambda -2}}{\left[ \left( \frac{k_T}{a} \right)^2 + \beta^2 \right]^{\delta/2}} \notag\\
    &\tag*{(substitute $s = k_T/a\beta$)} \notag\\
    &\hspace{-70pt}= \frac{2 \pi^{\left( d - 1 \right) / 2}}{\Gamma \! \left( \frac{d - 1}{2} \right)} a^{d+\lambda-1} \int_{-\infty}^{\infty} dk_d \left( k_d \right)^\kappa \, \beta^{\, d+\lambda-\delta-1} \int_{0}^{\infty} ds \, \frac{s^{d+\lambda-2}}{\left( s^2 + 1 \right) ^{\delta/2} } \notag\\
    &\tag*{(substitute $t = k_d/b m$)} \notag\\
    &\hspace{-70pt}= \frac{2 \pi^{\left( d - 1 \right) / 2}}{\Gamma \! \left( \frac{d - 1}{2} \right)} a^{d+\lambda-1} \, b^{\kappa+1} \,  m^{d+\kappa+\lambda-\delta} 
    \int_{-\infty}^{\infty} dt \ t^\kappa \left( t^2 + 1 \right) ^{\left( d+\lambda-\delta-1 \right) / 2} \int_{0}^{\infty} ds \, \frac{s^{d+\lambda-2}}{\left( s^2 + 1 \right) ^{\delta/2} }
\end{align}
The $s$ and $t$ integrals converge only under certain conditions. If $1 < d + \lambda < 1 + \delta$,
\begin{align}
    \int_{0}^{\infty} ds \, \frac{s^{d+\lambda-2}}{\left( s^2 + 1 \right) ^{\delta/2} } &= \frac{\Gamma \! \left( -\frac{d+\lambda-\delta-1}{2}\right) \Gamma \! \left( \frac{d+\lambda-1}{2}\right)}{2 \ \Gamma \! \left( \frac{\delta}{2}\right)}
\end{align}
If $d + \lambda <  \delta - \kappa \, , \ \kappa > -1$,
\begin{align}
    \int_{-\infty}^{\infty} dt \ t^\kappa \left( t^2 + 1 \right) ^{\left( d+\lambda-\delta-1 \right) / 2} &= \left( \frac{1+(-1)^\kappa}{2} \right) \frac{\Gamma \! \left( \frac{\kappa+1}{2}\right) \Gamma \! \left( -\frac{d+\kappa+\lambda-\delta}{2}\right)}{\Gamma \! \left( - \frac{d+\lambda-\delta-1}{2}\right)}
\end{align}
Analytically continuing these integrals,
\begin{align}
    P_{\,\rm{NC}} (d,\kappa,\lambda,\delta) &= \pi^{(d-1)/2} \left( \frac{1+(-1)^\kappa}{2} \right) \frac{\Gamma \! \left( \frac{\kappa+1}{2}\right) \Gamma \! \left( -\frac{d+\kappa+\lambda-\delta}{2}\right) \Gamma \! \left( \frac{d+\lambda-1}{2}\right)}{\Gamma \! \left( \frac{d-1}{2}\right) \Gamma \! \left( \frac{\delta}{2}\right)} \notag\\
    &\hspace{4mm}\times a^{d+\lambda-1} b^{\kappa+1} \, m^{d+\kappa+\lambda-\delta}
\end{align}

\subsection{Compact} \label{sec:PCAppendix}
\begin{align}
    P_{\rm{C}} ( d, \kappa, \lambda, \delta ) :\!&= \frac{1}{r_0} \sum_{n = -\infty}^{\infty} \int_{-\infty}^{\infty} \, \prod_{i=1}^{d-1} dk_i \, \frac{\left( n/r_0 \right) ^\kappa \left( k_T \right)^\lambda}{\left[ \left( \frac{k_T}{a} \right)^2 + \left( \frac{n/r_0}{b} \right)^2 + m^2 \right]^{\delta/2}} \notag\\
    &\tag*{(similar steps as the non-compact case)} \notag\\
    &\hspace{-45pt}= \frac{1}{r_0} \, \pi^{\left( d - 1 \right) / 2} \ \frac{\Gamma \! \left( -\frac{d+\lambda-\delta-1}{2}\right) \Gamma \! \left( \frac{d+\lambda-1}{2}\right)}{\Gamma \! \left( \frac{d - 1}{2} \right) \Gamma \! \left( \frac{\delta}{2} \right)} a^{d+\lambda-1} \sum_{n=-\infty}^{\infty} \left( \frac{n}{r_0} \right)^\kappa \left[ \left( \frac{n/r_0}{b} \right)^2 + m^2 \right]^\frac{d+\lambda-\delta-1}{2} \notag\\
    &\hspace{-45pt}= \pi^{\left( d - 1 \right) / 2} \ \frac{\Gamma \! \left( -\frac{d+\lambda-\delta-1}{2}\right) \Gamma \! \left( \frac{d+\lambda-1}{2}\right)}{\Gamma \! \left( \frac{d - 1}{2} \right) \Gamma \! \left( \frac{\delta}{2} \right)} a^{d+\lambda-1} \, r_{0}^{-d-\kappa-\lambda+\delta} \, b^{-d-\lambda+\delta+1} \, S( d, \kappa, \lambda, \delta )
\end{align}
where
\begin{align}
    S ( d, \kappa, \lambda, \delta ) : \! &= \sum_{n=-\infty}^{\infty} n^\kappa \left[ n^2 + \left( r_0 m b \right)^2 \right]^{\frac{d+\lambda-\delta-1}{2}} \notag\\
    &\tag*{(use the binomial theorem)} \notag\\
    &= \sum_{n=-\infty}^{\infty} \, \sum_{p=0}^{\kappa/2} \binom{\kappa/2}{p} \left[ n^2 + \left( r_0 m b \right)^2 \right]^{\frac{\kappa}{2} - p} \left[ -\left( r_0 m b \right)^{2} \right]^{p} \left[ n^2 + \left( r_0 m b \right)^2 \right]^{\frac{d+\lambda-\delta-1}{2}} \notag\\
    &= \sum_{p=0}^{\kappa/2} \left( -1 \right)^p \binom{\kappa/2}{p} \left( r_0 m b \right)^{2p} \zeta \! \left(s = p - \frac{\kappa}{2} - \frac{d+\lambda-\delta-1}{2}, \ \alpha = r_0 m b \right)\\
    \zeta ( s, \alpha ) : \! &= \sum_{n=-\infty}^{\infty} \left( n^2 + \alpha^2 \right)^{-s}
\end{align}

\section{Values of the \texorpdfstring{$\bm{P_{\rm{reg}}(d, \kappa, \lambda, \delta)}$}{} function} \label{sec:PRegValsappendix}

\subsection{\texorpdfstring{$d = 3$}{}} \label{sec:PRegVals3dappendix}

\renewcommand{\arraystretch}{1.5}

\begin{center}
\begin{tabular}{|m{1em}|m{1em}|m{1em}|m{31.7em}|} 
\hline
 $\kappa$ & $\lambda$ & $\delta$ & $P_{\rm{reg}} ( 3, \kappa, \lambda, \delta )$ \\ [0.5ex] 
 \hline
 0 & 0 & -1 & $-2 \pi a^2 b m^4 Q \! \left( -\frac{3}{2}, r_0 m b \right)$ \\
 \hline
 0 & 0 & 1 & $4 \pi a^2 b m^2 Q \! \left( -\frac{1}{2}, r_0 m b \right)$\\
 \hline
 0 & 2 & 1 & $4 \pi a^4 b m^4 Q \! \left( -\frac{3}{2}, r_0 m b \right)$\\
 \hline
 2 & 0 & 1 & $-2 \pi a^2 b^3 m^4 \left[ 3 \, Q \! \left( -\frac{3}{2}, r_0 m b \right) + 2\, Q \! \left( -\frac{1}{2}, r_0 m b \right) \right]$\\
 \hline
 0 & 2 & 3 & $8 \pi a^4 b m^2 Q \! \left( -\frac{1}{2}, r_0 m b \right)$ \\
 \hline
\end{tabular}
\end{center}

\begin{center}
\begin{tabular}{|m{1em}|m{1em}|m{1em}|m{31.7em}|} 
\hline
 $\kappa$ & $\lambda$ & $\delta$ & $P_{\rm{reg}} ( 3, \kappa, \lambda, \delta )$ \\ [0.5ex] 
 \hline
 2 & 0 & 3 & $- 4 \pi a^2 b^3 m^2 \left[ Q \! \left( -\frac{1}{2}, r_0 m b \right) + 2 \, Q \! \left( \frac{1}{2}, r_0 m b \right) \right]$ \\
 \hline
 0 & 4 & 5 & $\frac{32}{3} \pi a^6 b m^2 Q \! \left( -\frac{1}{2}, r_0 m b \right)$ \\
 \hline
 2 & 2 & 5 & $-\frac{8}{3} \pi a^4 b^3 m^2 \left[ Q \! \left( -\frac{1}{2}, r_0 m b \right) + 2 \, Q \! \left( \frac{1}{2}, r_0 m b \right) \right]$ \\
 \hline
 4 & 0 & 5 & $-\frac{4}{3} \pi a^2 b^5 m^2 \left[ Q \! \left( -\frac{1}{2}, r_0 m b \right) + 4 \, Q \! \left( \frac{1}{2}, r_0 m b \right) - 4 \, Q \! \left( \frac{3}{2}, r_0 m b \right) \right]$ \\
 \hline
 0 & 6 & 7 & $\frac{64}{5} \pi a^8 b m^2 Q \! \left( -\frac{1}{2}, r_0 m b \right)$\\
 \hline
 2 & 4 & 7 & $-\frac{32}{15} \pi a^6 b^3 m^2 \left[Q \! \left( -\frac{1}{2}, r_0 m b \right) + 2 \, Q \! \left( \frac{1}{2}, r_0 m b \right) \right]$\\
 \hline
 4 & 2 & 7 & $-\frac{8}{15} \pi a^4 b^5 m^2 \left[ Q \! \left( -\frac{1}{2}, r_0 m b \right) + 4 \, Q \! \left( \frac{1}{2}, r_0 m b \right) - 4 \, Q \! \left( \frac{3}{2}, r_0 m b \right) \right]$\\
 \hline
 6 & 0 & 7 & $-\frac{4}{15} \pi a^2 b^7 m^2 \left[ 3 \, Q \! \left( -\frac{1}{2}, r_0 m b \right) + 18 \, Q \! \left( \frac{1}{2}, r_0 m b \right) - 36 \, Q \! \left( \frac{3}{2}, r_0 m b \right) \right.$ \hspace*{25mm}$\left. + 8 \, Q \! \left( \frac{5}{2}, r_0 m b \right) \right]$\\
 \hline
 0 & 0 & 3 & $8 \pi a^2 b \, Q \! \left( \frac{1}{2}, r_0 m b \right)$ \\
 \hline
 0 & 2 & 5 & $\frac{16}{3} \pi a^4 b \, Q \! \left( \frac{1}{2}, r_0 m b \right)$ \\
 \hline
 2 & 0 & 5 & $\frac{8}{3} \pi a^2 b^3 \left[ Q \! \left( \frac{1}{2}, r_0 m b \right) - 2 \, Q \! \left( \frac{3}{2}, r_0 m b \right) \right]$ \\ 
 \hline
 0 & 4 & 7 & $\frac{64}{15} \pi a^6 b \, Q \! \left( \frac{1}{2}, r_0 m b \right)$ \\
 \hline
 2 & 2 & 7 & $\frac{16}{15} \pi a^4 b^3 \left[ Q \! \left( \frac{1}{2}, r_0 m b \right) - 2 \, Q \! \left( \frac{3}{2}, r_0 m b \right) \right]$ \\
 \hline
 4 & 0 & 7 & $\frac{8}{15} \pi a^2 b^5 \left[ 3 \, Q \! \left( \frac{1}{2}, r_0 m b \right) - 12 \, Q \! \left( \frac{3}{2}, r_0 m b \right) + 4 \, Q \! \left( \frac{5}{2}, r_0 m b \right)\right]$ \\
 \hline
 0 & 6 & 9 & $\frac{128}{35} \pi a^8 b \, Q \! \left( \frac{1}{2}, r_0 m b \right)$ \\ 
 \hline
 2 & 4 & 9 & $\frac{64}{105} \pi a^6 b^3 \left[ Q \! \left( \frac{1}{2}, r_0 m b \right) - 2 \, Q \! \left( \frac{3}{2}, r_0 m b \right) \right]$ \\
 \hline
 4 & 2 & 9 & $\frac{16}{105} \pi a^4 b^5 \left[ 3 \, Q \! \left( \frac{1}{2}, r_0 m b \right) - 12 \, Q \! \left( \frac{3}{2}, r_0 m b \right) + 4 \, Q \! \left( \frac{5}{2}, r_0 m b \right)\right]$ \\ 
 \hline
 6 & 0 & 9 & $\frac{8}{105} \pi a^2 b^7 \left[ 15 \, Q \! \left( \frac{1}{2}, r_0 m b \right) - 90 \, Q \! \left( \frac{3}{2}, r_0 m b \right) + 60 \, Q \! \left( \frac{5}{2}, r_0 m b \right) \right.$ \hspace*{25mm}$\left. - 8 \, Q \! \left( \frac{7}{2}, r_0 m b \right)\right]$ \\
 \hline
 0 & 8 & 11 & $\frac{1024}{315} \pi a^{10} b \, Q \! \left( \frac{1}{2}, r_0 m b \right)$ \\
 \hline
 2 & 6 & 11 & $\frac{128}{315} \pi a^8 b^3 \, \left[ Q \! \left( \frac{1}{2}, r_0 m b \right) - 2 \, Q \! \left( \frac{3}{2}, r_0 m b \right) \right]$ \\
 \hline
 4 & 4 & 11 & $\frac{64}{945} \pi a^6 b^5 \left[ 3 \, Q \! \left( \frac{1}{2}, r_0 m b \right) - 12 \, Q \! \left( \frac{3}{2}, r_0 m b \right) + 4 \, Q \! \left( \frac{5}{2}, r_0 m b \right)\right]$ \\
 \hline
 6 & 2 & 11 & $\frac{16}{945} \pi a^4 b^7 \left[ 15 \, Q \! \left( \frac{1}{2}, r_0 m b \right) - 90 \, Q \! \left( \frac{3}{2}, r_0 m b \right) + 60 \, Q \! \left( \frac{5}{2}, r_0 m b \right) \right.$ \hspace*{25mm}$\left. - 8 \, Q \! \left( \frac{7}{2}, r_0 m b \right)\right]$ \\
 \hline
 8 & 0 & 11 & $\frac{8}{945} \pi a^2 b^9 \left[ 105 \, Q \! \left( \frac{1}{2}, r_0 m b \right) - 840 \, Q \! \left( \frac{3}{2}, r_0 m b \right) + 840 \, Q \! \left( \frac{5}{2}, r_0 m b \right) \right.$ \hspace*{25mm}$\left. - 224 \, Q \! \left( \frac{7}{2}, r_0 m b \right) + 16 \, Q \! \left( \frac{9}{2}, r_0 m b \right) \right]$ \\
 \hline
 0 & 10 & 13 & $\frac{2048}{693} \pi a^{12} b \, Q \! \left( \frac{1}{2}, r_0 m b \right)$ \\
 \hline
 2 & 8 & 13 & $\frac{1024}{3465} \pi a^{10} b^3 \left[ Q \! \left( \frac{1}{2}, r_0 m b \right) - 2 \, Q \! \left( \frac{3}{2}, r_0 m b \right) \right]$\\
 \hline
\end{tabular}
\end{center}

\begin{center}
\begin{tabular}{|m{1em}|m{1em}|m{1em}|m{31.7em}|} 
\hline
 $\kappa$ & $\lambda$ & $\delta$ & $P_{\rm{reg}} ( 3, \kappa, \lambda, \delta )$ \\ [0.5ex] 
 \hline
 4 & 6 & 13 & $\frac{128}{3465} \pi a^8 b^5 \left[ 3 \, Q \! \left( \frac{1}{2}, r_0 m b \right) - 12 \, Q \! \left( \frac{3}{2}, r_0 m b \right) + 4 \, Q \! \left( \frac{5}{2}, r_0 m b \right)\right]$\\
 \hline
 6 & 4 & 13 & $\frac{64}{10395} \pi a^6 b^7 \left[ 15 \, Q \! \left( \frac{1}{2}, r_0 m b \right) - 90 \, Q \! \left( \frac{3}{2}, r_0 m b \right) + 60 \, Q \! \left( \frac{5}{2}, r_0 m b \right) \right.$ \hspace*{25mm}$\left. - 8 \, Q \! \left( \frac{7}{2}, r_0 m b \right)\right]$\\
 \hline
 8 & 2 & 13 & $\frac{16}{10395} \pi a^4 b^9 \left[ 105 \, Q \! \left( \frac{1}{2}, r_0 m b \right) - 840 \, Q \! \left( \frac{3}{2}, r_0 m b \right) + 840 \, Q \! \left( \frac{5}{2}, r_0 m b \right) \right.$ \hspace*{25mm}$\left. - 224 \, Q \! \left( \frac{7}{2}, r_0 m b \right) + 16 \, Q \! \left( \frac{9}{2}, r_0 m b \right) \right]$ \\
 \hline
 10 & 0 & 13 & $\frac{8}{10395} \pi a^2 b^{11} \left[ 945 \, Q \! \left( \frac{1}{2}, r_0 m b \right) - 9450 \, Q \! \left( \frac{3}{2}, r_0 m b \right) + 12600 \, Q \! \left( \frac{5}{2}, r_0 m b \right) \right.$ \hspace*{25mm}$\left. - 5040 \, Q \! \left( \frac{7}{2}, r_0 m b \right) + 720 \, Q \! \left( \frac{9}{2}, r_0 m b \right) - 32 \, Q \! \left( \frac{11}{2}, r_0 m b \right) \right]$\\
 \hline
\end{tabular}
\end{center}

\subsection{\texorpdfstring{$d = 4$}{}} \label{sec:PRegVals4dappendix}

\renewcommand{\arraystretch}{1.5}

\begin{center}
\begin{tabular}{|m{1em}|m{1em}|m{1em}|m{31.7em}|} 
\hline
 $\kappa$ & $\lambda$ & $\delta$ & $P_{\rm{reg}} ( 4, \kappa, \lambda, \delta )$ \\ [0.5ex] 
 \hline
 0 & 0 & -1 & $-2 \pi^{3/2} a^3 b m^5 Q \! \left( -2, r_0 m b \right)$ \\
 \hline
 0 & 0 & 1 & $4 \pi^{3/2} a^3 b m^3 Q \! \left( -1, r_0 m b \right)$\\
 \hline
 0 & 2 & 1 & $6 \pi^{3/2} a^5 b m^5 Q \! \left( -2, r_0 m b \right)$\\
 \hline
 2 & 0 & 1 & $-4 \pi^{3/2} a^3 b^3 m^5 \left[ 2 \, Q \! \left( -2, r_0 m b \right) + Q \! \left( -1, r_0 m b \right) \right]$\\
 \hline
 0 & 2 & 3 & $12 \pi^{3/2} a^5 b m^3 Q \! \left( -1, r_0 m b \right)$ \\
 \hline
 2 & 0 & 3 & $- 8 \pi^{3/2} a^3 b^3 m^3 \left[ Q \! \left( -1, r_0 m b \right) + Q \! \left( 0, r_0 m b \right) \right]$ \\
 \hline
 0 & 4 & 5 & $20 \pi^{3/2} a^7 b m^3 Q \! \left( -1, r_0 m b \right)$ \\
 \hline
 2 & 2 & 5 & $- 8 \pi^{3/2} a^5 b^3 m^3 \left[ Q \! \left( -1, r_0 m b \right) + Q \! \left( 0, r_0 m b \right) \right]$ \\
 \hline
 4 & 0 & 5 & $\frac{16}{3} \pi^{3/2} a^3 b^5 m^3 Q \! \left( 1, r_0 m b \right)$ \\
 \hline
 0 & 6 & 7 & $28 \pi^{3/2} a^9 b m^3 Q \! \left( -1, r_0 m b \right)$\\
 \hline
 2 & 4 & 7 & $-8 \pi^{3/2} a^7 b^3 m^3 \left[Q \! \left( -1, r_0 m b \right) + Q \! \left( 0, r_0 m b \right) \right]$\\
 \hline
 4 & 2 & 7 & $\frac{16}{5} \pi^{3/2} a^5 b^5 m^3 Q \! \left( 1, r_0 m b \right)$\\
 \hline
 6 & 0 & 7 & $\frac{32}{15} \pi^{3/2} a^3 b^7 m^3 \left[3 \, Q \! \left( 1, r_0 m b \right) - Q \! \left( 2, r_0 m b \right) \right]$\\
 \hline
 0 & 0 & 3 & $8 \pi^{3/2} a^3 b m \, Q \! \left( 0, r_0 m b \right)$ \\
 \hline
 0 & 2 & 5 & $8 \pi^{3/2} a^5 b m \, Q \! \left( 0, r_0 m b \right)$ \\
 \hline
 2 & 0 & 5 & $- \frac{16}{3} \pi^{3/2} a^3 b^3 m \, Q \! \left( 1, r_0 m b \right)$ \\ 
 \hline
 0 & 4 & 7 & $8 \pi^{3/2} a^7 b m \, Q \! \left( 0, r_0 m b \right)$ \\
 \hline
 2 & 2 & 7 & $- \frac{16}{5} \pi^{3/2} a^5 b^3 m \, Q \! \left( 1, r_0 m b \right)$ \\
 \hline
\end{tabular}
\end{center}

\begin{center}
\begin{tabular}{|m{1em}|m{1em}|m{1em}|m{31.7em}|} 
\hline
 $\kappa$ & $\lambda$ & $\delta$ & $P_{\rm{reg}} ( 4, \kappa, \lambda, \delta )$ \\ [0.5ex] 
 \hline
 4 & 0 & 7 & $- \frac{32}{15} \pi^{3/2} a^3 b^5 m \left[ 2 \, Q \! \left( 1, r_0 m b \right) - Q \! \left( 2, r_0 m b \right) \right]$ \\
 \hline
 0 & 6 & 9 & $8 \pi^{3/2} a^9 b m \, Q \! \left( 0, r_0 m b \right)$ \\ 
 \hline
 2 & 4 & 9 & $- \frac{16}{7} \pi^{3/2} a^7 b^3 m \, Q \! \left( 1, r_0 m b \right)$ \\
 \hline
 4 & 2 & 9 & $- \frac{32}{35} \pi^{3/2} a^5 b^5 m \left[ 2 \, Q \! \left( 1, r_0 m b \right) - Q \! \left( 2, r_0 m b \right) \right]$ \\ 
 \hline
 6 & 0 & 9 & $- \frac{64}{105} \pi^{3/2} a^3 b^7 m \left[ 6 \, Q \! \left( 1, r_0 m b \right) - 6 \, Q \! \left( 2, r_0 m b \right) + Q \! \left( 3, r_0 m b \right) \right]$ \\
 \hline
 0 & 8 & 11 & $8 \pi^{3/2} a^{11} b m \, Q \! \left( 0, r_0 m b \right)$ \\
 \hline
 2 & 6 & 11 & $- \frac{16}{9} \pi^{3/2} a^9 b^3 m \, Q \! \left( 1, r_0 m b \right)$ \\
 \hline
 4 & 4 & 11 & $- \frac{32}{63} \pi^{3/2} a^7 b^5 m \left[ 2 \, Q \! \left( 1, r_0 m b \right) - Q \! \left( 2, r_0 m b \right) \right]$ \\
 \hline
 6 & 2 & 11 & $- \frac{64}{315} \pi^{3/2} a^5 b^7 m \left[ 6 \, Q \! \left( 1, r_0 m b \right) - 6 \, Q \! \left( 2, r_0 m b \right) + Q \! \left( 3, r_0 m b \right) \right]$ \\
 \hline
 8 & 0 & 11 & $- \frac{128}{945} \pi^{3/2} a^3 b^9 m \left[ 24 \, Q \! \left( 1, r_0 m b \right) - 36 \, Q \! \left( 2, r_0 m b \right) + 12 \, Q \! \left( 3, r_0 m b \right) \right.$ \hspace*{30mm}$\left. - Q \! \left( 4, r_0 m b \right) \right]$ \\
 \hline
 0 & 10 & 13 & $8 \pi^{3/2} a^{13} b m \, Q \! \left( 0, r_0 m b \right)$ \\
 \hline
 2 & 8 & 13 & $-\frac{16}{11} \pi^{3/2} a^{11} b^3 m \, Q \! \left( 1, r_0 m b \right)$\\
 \hline
 4 & 6 & 13 & $-\frac{32}{99} \pi^{3/2} a^9 b^5 m \left[ 2 \, Q \! \left( 1, r_0 m b \right) - Q \! \left( 2, r_0 m b \right) \right]$\\
 \hline
 6 & 4 & 13 & $-\frac{64}{693} \pi^{3/2} a^7 b^7 m \left[ 6 \, Q \! \left( 1, r_0 m b \right) - 6 \, Q \! \left( 2, r_0 m b \right) + Q \! \left( 3, r_0 m b \right) \right]$\\
 \hline
 8 & 2 & 13 & $-\frac{128}{3465} \pi^{3/2} a^5 b^9 m \left[ 24 \, Q \! \left( 1, r_0 m b \right) - 36 \, Q \! \left( 2, r_0 m b \right) + 12 \, Q \! \left( 3, r_0 m b \right) \right.$ \hspace*{30mm}$\left. - Q \! \left( 4, r_0 m b \right) \right]$ \\
 \hline
 10 & 0 & 13 & $-\frac{256}{10395} \pi^{3/2} a^3 b^{11} m \left[ 120 \, Q \! \left( 1, r_0 m b \right) - 240 \, Q \! \left( 2, r_0 m b \right) + 120 \, Q \! \left( 3, r_0 m b \right) \right.$  \hspace*{30mm}       $\left. - 20 \, Q \! \left( 4, r_0 m b \right) + Q \! \left( 5, r_0 m b \right) \right]$\\
 \hline
\end{tabular}
\end{center}

\section{Stress-energy expectation in the conformally flat scenario} \label{sec:AdExactValsappendix}

For the conformally flat scenario $a(t) = b(t)$, the metric \eqref{eq:d_FLRW} can be expressed in conformal coordinates as 
\begin{align}
    ds^2 = a(\eta)^2 \left( d\eta^2 - \sum_{j=1}^{d} dx_j^2 \right)
    \label{5dconformalFLRW}
\end{align}
where $\eta(t) = \int^t_{t_0} \frac{1}{a(t')} \, dt'$ is the conformal time. In the massless, conformally coupled limit ($m \to 0, \, \xi \to \frac{d-1}{4 d}$), the field equation of motion 
\begin{align}
    \left( \Box + \frac{d-1}{4 d} \, \mathcal{R}_{(1+d)} \right) \Phi = 0
    \label{genericconformalEOM}
\end{align}
is solvable with exact modes of the form
\begin{align}
    f_{nk}(x) &= \left( 2 ( 2 \pi )^{d-1} \left( 2 \pi r_0 \right) \right)^{-1/2} a(\eta)^{-(d-1)/2} e^{i \left( \vec{k} \cdot \vec{x} + \frac{n}{r_0} x_d \right) } \, h_{nk}(\eta) \\
    h_{nk}(\eta) &= \frac{1}{\sqrt{\omega_{nk}}} \, e^{-i \omega_{nk} \eta}, \hspace{5mm} \omega_{nk} = \sqrt{\sum_{i=1}^{d-1} k_i^2 + \left(\frac{n}{r_0} \right)^2}
\end{align}
Using these exact mode solutions, the standard adiabatic regularization prescription may be used to compute the stress-energy expectation. The results for $d=3$ and $d=4$ are listed below. Note that the scale factor $a = a(\eta)$ is a function of the conformal time, and all the derivatives in this section are with respect to $\eta$. It can be checked that the expressions below match (after performing the appropriate time coordinate transformation) the results obtained using the modified prescription in sections~\ref{sec:d3Scenario} and \ref{sec:d4Scenario}.

\subsection{\texorpdfstring{$d = 3$}{}} \label{sec:AdExactVals3dappendix}

\begin{align}
    \vev{T_{\eta \eta}}_{\rm{reg}}^{\left( 0 \right)} =& -\frac{1}{1440 \, \pi^2 \, r_0^4 \, a^2}\\
    \vev{T_{ii}}_{\rm{reg}}^{\left( 0 \right)} =& \, \frac{1}{1440 \, \pi^2 \, r_0^4 \, a^2}\\
    \vev{T_{33}}_{\rm{reg}}^{\left( 0 \right)} =& -\frac{1}{480 \, \pi^2 \, r_0^4 \, a^2}\\
    \vev{T_{\eta \eta}}_{\rm{reg}}^{\left( 2 \right)} =& \, \vev{T_{ii}}_{\rm{reg}}^{\left( 2 \right)} = \vev{T_{33}}_{\rm{reg}}^{\left( 2 \right)} = 0\\
    \vev{T_{\eta \eta}}_{\rm{reg}}^{\left( 4 \right)} =& \, \frac{\left(a'\right)^4}{960 \pi ^2 a^6}-\frac{\left(a''\right)^2}{960 \pi ^2 a^4}-\frac{\left(a'\right)^2 a''}{240 \pi ^2 a^5}+\frac{a^{(3)} a'}{480 \pi ^2 a^4}\\
    \vev{T_{ii}}_{\rm{reg}}^{\left( 4 \right)} =& \, \vev{T_{33}}_{\rm{reg}}^{\left( 4 \right)} = \frac{\left(a'\right)^4}{576 \pi ^2 a^6}-\frac{a^{(4)}}{1440 \pi ^2 a^3}+\frac{\left(a''\right)^2}{576 \pi ^2 a^4}-\frac{\left(a'\right)^2 a''}{144 \pi ^2 a^5}+\frac{a^{(3)} a'}{288 \pi ^2 a^4}
\end{align}

\subsection{\texorpdfstring{$d = 4$}{}} \label{sec:AdExactVals4dappendix}

\begin{align}
    \vev{T_{\eta \eta}}_{\rm{reg}}^{\left( 0 \right)} =& -\frac{3 \, \zeta (5)}{128 \, \pi^7 \, r_0^5 \, a^3}\\
    \vev{T_{ii}}_{\rm{reg}}^{\left( 0 \right)} =& \, \frac{3 \, \zeta (5)}{128 \, \pi^7 \, r_0^5 \, a^3}\\
    \vev{T_{44}}_{\rm{reg}}^{\left( 0 \right)} =& -\frac{3 \, \zeta (5)}{32 \, \pi^7 \, r_0^5 \, a^3}\\
    \vev{T_{\eta \eta}}_{\rm{reg}}^{\left( 2 \right)} =& \, \vev{T_{ii}}_{\rm{reg}}^{\left( 2 \right)} = \vev{T_{44}}_{\rm{reg}}^{\left( 2 \right)} = 0\\
    \vev{T_{\eta \eta}}_{\rm{reg}}^{\left( 4 \right)} =& \, \vev{T_{ii}}_{\rm{reg}}^{\left( 4 \right)} = \vev{T_{44}}_{\rm{reg}}^{\left( 4 \right)} = 0
\end{align}

\section{Right-hand sides of equations \texorpdfstring{\eqref{eq:einframeeq1}-\eqref{eq:einframeeq3}}{}} \label{sec:RHSEFE}

\begin{align}
    T_1^{(0)} =& -\frac{3 \, \zeta (5)}{64 \, \pi^6 \, r_0^4 \, M_4^2} e^{-2 \sqrt{6} \chi/M_4}\\
    T_1^{(2)} =& \, \frac{\zeta (3)}{128 \, \pi^4 \, r_0^2 \, M_4^2} e^{-\sqrt{6} \chi/M_4} \left(-\sqrt{\frac{3}{2}} \frac{c'}{c} \frac{\chi'}{M_4} + \frac{1}{2} \frac{(c')^2}{c^2} + \frac{3}{4} \frac{(\chi')^2}{M_4^2}\right)\\
    T_1^{(4)} =& \, \frac{3 \gamma }{2048 \, \pi ^2 \, M_4^2} \Biggl( \sqrt{\frac{2}{3}} \frac{c^{(3)}}{c} \frac{\chi'}{M_4} - \sqrt{\frac{2}{3}} \frac{c''}{c} \frac{\chi''}{M_4} - \frac{c''}{c} \frac{(\chi')^2}{M_4^2} + \frac{1}{3} \frac{(c'')^2}{c^2} + \sqrt{\frac{2}{3}} \frac{c'}{c} \frac{\chi^{(3)}}{M_4} \notag\\
    &- 3\sqrt{6} \, \frac{c'}{c} \frac{(\chi')^3}{M_4^3} - 2 \, \frac{c'}{c} \frac{\chi' \chi''}{M_4^2} + \frac{(c')^4}{c^4} - 5\sqrt{\frac{2}{3}} \frac{(c')^3}{c^3} \frac{\chi'}{M_4} + \sqrt{6} \, \frac{(c')^2}{c^2} \frac{\chi''}{M_4} + \sqrt{\frac{2}{3}} \frac{c' c''}{c^2} \frac{\chi'}{M_4} \notag\\
    &+ \frac{17}{2} \frac{(c')^2}{c^2} \frac{(\chi')^2}{M_4^2} - \frac{2}{3} \frac{(c')^2 c''}{c^3} - \frac{2}{3} \frac{c^{(3)} c'}{c^2} + \frac{1}{2} \frac{(\chi'')^2}{M_4^2} + \frac{9}{4} \frac{(\chi')^4}{M_4^4} - \frac{\chi^{(3)} \chi'}{M_4^2} \Biggr)\\
    T_2^{(0)} =& \, \frac{3 \, \zeta (5)}{64 \, \pi^6 \, r_0^4 \, M_4^2} e^{-2 \sqrt{6} \chi/M_4}\\
    T_2^{(2)} =& \, \frac{\zeta (3)}{128 \, \pi^4 \, r_0^2 \, M_4^2} e^{-\sqrt{6} \chi/M_4} \left(-\frac{1}{3} \frac{c''}{c} + \sqrt{\frac{2}{3}} \frac{c'}{c} \frac{\chi'}{M_4} - \frac{1}{6} \frac{(c')^2}{c^2} + \frac{1}{\sqrt{6}} \frac{\chi''}{M_4} - \frac{1}{4} \frac{(\chi')^2}{M_4^2}\right)\\
    T_2^{(4)} =& \, \frac{\gamma }{2048 \, \pi ^2 \, M_4^2} \Biggl( \frac{2}{3} \frac{c^{(4)}}{c} - 5\sqrt{\frac{2}{3}} \frac{c''}{c} \frac{\chi''}{M_4} - 7 \, \frac{c''}{c} \frac{(\chi')^2}{M_4^2} + \frac{(c'')^2}{c^2} - 5\sqrt{\frac{2}{3}} \frac{c'}{c} \frac{\chi^{(3)}}{M_4} - 14 \, \frac{c'}{c} \frac{\chi' \chi''}{M_4^2} \notag\\
    &+ \frac{(c')^4}{c^4} + 2\sqrt{\frac{2}{3}} \frac{(c')^2}{c^2} \frac{\chi''}{M_4} + 3\sqrt{6} \, \frac{c' c''}{c^2} \frac{\chi'}{M_4} - \frac{7}{2} \frac{(c')^2}{c^2} \frac{(\chi')^2}{M_4^2} - 4 \, \frac{(c')^2 c''}{c^3} + \frac{4}{3} \frac{c^{(3)} c'}{c^2} \notag\\
    &- \sqrt{\frac{2}{3}} \frac{\chi^{(4)}}{M_4} + \frac{1}{2} \frac{(\chi'')^2}{M_4^2} + \frac{9}{4} \frac{(\chi')^4}{M_4^4} - \frac{\chi^{(3)} \chi'}{M_4^2} + 3\sqrt{6} \, \frac{(\chi')^2 \chi''}{M_4^3} \Biggr)\\
    T_3^{(0)} =& -\frac{3 \, \zeta (5)}{16 \, \pi^6 \, r_0^4 \, M_4^2} e^{-2 \sqrt{6} \chi/M_4}\\
    T_3^{(2)} =& \, \frac{\zeta (3)}{128 \, \pi^4 \, r_0^2 \, M_4^2} e^{-\sqrt{6} \chi/M_4} \left(\frac{c''}{c} - 3\sqrt{\frac{3}{2}} \frac{c'}{c} \frac{\chi'}{M_4} + \frac{(c')^2}{c^2} - \sqrt{\frac{3}{2}} \frac{\chi''}{M_4} + \frac{3}{2} \frac{(\chi')^2}{M_4^2}\right)\\
    T_3^{(4)} =& \, \frac{\gamma }{1024 \, \pi ^2 \, M_4^2} \Biggl( -\frac{c^{(4)}}{c} + \sqrt{\frac{3}{2}} \frac{c^{(3)}}{c} \frac{\chi'}{M_4} + 2\sqrt{6} \, \frac{c''}{c} \frac{\chi''}{M_4} + 9 \, \frac{c''}{c} \frac{(\chi')^2}{M_4^2} - \frac{(c'')^2}{c^2} + 3\sqrt{6} \, \frac{c'}{c} \frac{\chi^{(3)}}{M_4} \notag\\
    &- 9\sqrt{\frac{3}{2}} \frac{c'}{c} \frac{(\chi')^3}{M_4^3} + 18 \, \frac{c'}{c} \frac{\chi' \chi''}{M_4^2} - 5\sqrt{\frac{3}{2}} \frac{(c')^3}{c^3} \frac{\chi'}{M_4} + \sqrt{\frac{3}{2}} \frac{(c')^2}{c^2} \frac{\chi''}{M_4} - 4\sqrt{6} \, \frac{c' c''}{c^2} \frac{\chi'}{M_4} \notag\\
    &+ 18 \, \frac{(c')^2}{c^2} \frac{(\chi')^2}{M_4^2} + 5 \, \frac{(c')^2 c''}{c^3} - 3 \, \frac{c^{(3)} c'}{c^2} + \sqrt{\frac{3}{2}} \frac{\chi^{(4)}}{M_4} - 9\sqrt{\frac{3}{2}} \frac{(\chi')^2 \chi''}{M_4^3} \Biggr)
\end{align}

\end{appendices}

\bibliographystyle{JHEP}
\bibliography{biblio.bib}

\end{document}